\documentclass[modern]{aastex61}

\shorttitle{RADIO BURSTS IN THE 2017 SEPTEMBER 6 FLARE}
\shortauthors{Karlick\'y and Ryb\'ak}
\begin{document}

\title{THE 2017 SEPTEMBER 6 FLARE: RADIO BURSTS AND PULSATIONS IN THE 22-5000 MHz RANGE AND ASSOCIATED PHENOMENA}

\author[0000-0002-3963-8701]{Marian Karlick\'y}
\affil{Astronomical Institute of the Academy of Sciences of the Czech Republic,
  CZ-25165 Ond\v{r}ejov, Czech Republic}
\author[0000-0003-3128-8396]{J\'an Ryb\'ak}
\affil{Astronomical Institute, Slovak Academy of Sciences,
  SK-05960 Tatransk\'{a} Lomnica, Slovakia}

\correspondingauthor{J\'an Ryb\'ak} \email{rybak@astro.sk}

\begin{abstract}
    For the 2017 September 6 flare (SOL2017-Sep-06T11:53)
    we present not only unusual radio bursts, but also their interesting time
    association with the other flare phenomena observed in EUV, white-light,
    X-ray, and $\gamma$-ray emissions. Using our new method based on wavelets
    we found quasi-periodic pulsations (QPPs) in several locations of the whole time-frequency
    domain of the analyzed radio spectrum (11:55--12:07 UT and 22--5000 MHz).
    Among them the drifting QPPs are new and the most interesting, especially
    a bi-directional
    QPP at the time of the hard X-ray and $\gamma$-ray peaks, and a sunquake start.
    In the pre-impulsive phase we show an unusual drifting pulsation structure
    (DPS) in association with the EUV brightenings caused by the interaction of
    magnetic ropes. In the flare impulsive phase we found an exceptional radio burst drifting from 5000 MHz
    to 800 MHz. In connection with this drifting burst, we show a U-burst
    at about the onset time of an EUV writhed structure and a drifting radio burst as a signature of a shock wave
    at high frequencies (1050--1350 MHz).
    In the peak flare phase we found an indication of an additional energy-release process
    located at higher altitudes in the solar atmosphere.
    These phenomena are interpreted considering a rising magnetic rope, magnetosonic waves and particle beams.
    Using a density model
    we estimated the density, wave velocities and source heights
    for the bi-directionally drifting QPPs, the density for the pre-impulsive DPS and U-burst, and the density and magnetic field strength
    for the drifting radio burst.
\end{abstract}

\keywords{Sun: flares -- Sun: radio radiation -- Sun: oscillations}

\section{Introduction}

Radio bursts are an integral part of solar flare observations. These bursts
were classified into several types (II, III, V, J, U, and IV) and summarized in
several schemes in the time-frequency domain
\citep{1979itsr.book.....K,1985srph.book.....M,1986SoPh..104...19P}. They
indicate superthermal electrons in solar flares. Many model for their
explanation were proposed, see e.g. the books by
\cite{1980gbs..bookR....M,1985srph.book.....M}. A catalogue of radio
bursts in the 1000-3000 MHz range and their statistics in the 800-2000 MHz
range can be found in the papers by
\cite{1994A&AS..104..145I,2001A&A...375..243J}.

Quasi-periodic pulsations (QPPs) are also a common feature of solar flares.
They were detected in all energy bands, from radio up to $\gamma$-rays.
Relations between the QPPs in different bands are of high interest and they
were studied in many papers. For example, \cite{1990A&A...229..206A} showed a
weak correlation between decimetric QPPs and hard X-ray emission. On the other
hand, a good correlation between the radio on 3 GHz, X-ray (11-26 keV), and EUV
(18 nm) was found by \cite{2020ApJ...888...18K}. For an explanation of the QPPs
several models were suggested. \cite{1987SoPh..111..113A} classified these
models in three groups according to their driving mechanisms: 1) magnetic flux
tube oscillations, 2) cyclic self-organizing systems of plasma instabilities,
and 3) modulation of particle acceleration. Furthermore, in the paper by
\cite{2009SSRv..149..119N}, the authors split the QPP models into two groups:
"load/unload" mechanisms (e.g. the repetitive regimes of flaring energy
releases by magnetic reconnection) and magnetohydrodynamic oscillations. For
the repetitive regime of the magnetic reconnection, see also the paper by
\cite{2000A&A...360..715K}. A review of magnetohydrodynamic oscillatory
processes in the solar corona is presented in \cite{2016SSRv..200...75N}. As
concerns observational verifications of these models, it looks that the most
probable models are those based on quasi-periodic acceleration and injection of
fast electrons \citep{2008ApJ...684.1433F,2019ApJ...875...98M} or those based
on the modulation by magnetohydrodynamic
waves~\citep{2006A&A...452..343N,2019NatCo..10.2276C}.

An interesting example of radio QPPs was presented in the paper by
\cite{2017SoPh..292....1K}. Here, during the 2010 August 1 flare, QPPs driftingx
from 2000 to 400 MHz and with a period of 160-220 s were found.
These QPPs were attributed to a fast mode magnetosonic wave train with a 181 s period
propagating upwards in the solar atmosphere \citep{2011ApJ...736L..13L}.

In the present paper, based on the analysis of the 22--5000 MHz radio spectrum
of the 2017 September 6 flare and in comparison with EUV, white-light, X-ray,
and $\gamma$-ray observations, we show several flare induced phenomena not
described in literature so far. For example, the burst at the beginning of the
flare that drifts from 5000 to 800 MHz  (burst B, see the following) differs
from the bursts in the radio schemes shown in
\cite{1979itsr.book.....K,1985srph.book.....M,1986SoPh..104...19P}. Analyzing
the radio spectrum we searched for periods and phases of significant
quasi-periodic pulsations in the whole 22--5000 MHz radio spectrum using our
novel technique based on the wavelet transform
\citep{2017SoPh..292....1K,2017SoPh..292...94K}. We found pulsations not only
on single frequencies, but in the all time-frequency sub-domains where they are
statistically significant. This technique filters the radio spectrum in the
selected intervals of time variations and thus enables to show fine features
that are on the original spectrum  hardly distinguishable. By this way we
recognized bi-directionally drifting QPPs that were not found so far. Just the
analysis of these drifting QPPs is one of main topics of our paper. We also
present interesting time associations between the observed radio bursts and
phenomena shown in other papers about this flare. Therefore, our results are
complementary to those shown in these papers.

The paper is structured as follows. In Section 2 we present previous
observations relevant to our study and describe the 22--5000 MHz radio
spectrum of the 2017 September 6 flare. An analysis of the quasi-periodic
pulsations in this radio spectrum together with an association of the radio
bursts with the phenomena in EUV, white-light, X- and $\gamma$-ray observations
are given in Section 3. Interpretation and discussion of the presented results
are in Section 4 and conclusions are summarized in Section 5.

\section{Observations}

\subsection{Previous observations}

On 2017 September 6, in the active region NOAA AR12673, the largest flare in
Solar Cycle 24, classified as X9.3 flare, was observed. According to the
(\textit{GOES}) observations this flare started at 11:53 UT, reached its peak
at 12:02 UT, and ended at 12:10 UT. It was accompanied by a large CME
\citep{2018ApJ...856...79Y}. This flare was described and analyzed in many
papers so far. In one channel of the irradiance measurements the first peak
appeared already in the pre-impulsive phase at 11:53:20 UT
\citep{2018ApJ...867L..24D}. At the flare location \cite{2018A&A...619A.100H}
recognized the double-decker flux rope configuration by the non-linear force
free field technique. This configuration erupted as shown by the brightenings
starting from 11:53:53 UT and rising writhed structure at 11:56:08 UT. The
flare impulsive phase was associated with X-ray up to $\gamma$-ray emissions.
The 300-1200 keV time profile showed two peaks: the first stronger one at
11:56--11:56:50 UT and weaker one at 11:57--11:58 UT
\citep{2019ApJ...877..145L}. During the first peak the analysis revealed
contributions from nuclear deexcitation lines, electron - positron annihilation
line at 511 keV, and a neutron capture line at 2.223 MeV along with two
components of the bremsstrahlung continuum \citep{2019ApJ...877..145L}.
\cite{2020ApJ...888...53L} analyzed periods in this flare, in the hard X-rays,
$\gamma$-rays and in radio on 1250.9 MHz. In the hard X-rays and 1250.9 MHz
they found the periods 24--30 s in the time interval 11:57--11:58 UT, and in
the $\gamma$-rays the period $\sim$20 s at 11:55:30--11:57 UT. Moreover in this
flare, the white-light flare emission and a sunquake were observed
\citep{2018A&A...620A.183J,2018ApJ...864...86S}. Continuum intensity variations
were detected close to double-decker flux rope configuration starting at
11:54:49 UT and this location was found as that where the helioseismic wave
started \citep{2018ApJ...864...86S}.

\subsection{Data}
For preparation of the broadband 22-5000 MHz radio spectrum we used data from
four different radiospectrographs: Greenland-Callisto radiospectrograph working
in the 22--100 MHz range with the resolutions of 0.25 s and 0.48 MHz
\citep{2016JSARA..11...34M}, ORFEES radiospectrograph working in the 150--1000
MHz range with the resolutions of 1.0 s and 0.98 MHz (publicly available data,
Observation Radio Frequence pour l'Etude des Eruptions Solaires
radiospectrograph, Nancay, France), and the Ond\v{r}ejov radiospectrographs
working in the ranges 800--2000 MHz and 2000--5000 MHz with the resolutions of
0.01 s and 4.7 MHz, and 0.01 s and 11.7 MHz \citep{2008SoPh..253...95J},
respectively. At some frequencies, human-made radio interference affected the
acquired signal. Such parts of the data are excluded and they are marked by
dark bands in all plots of the radiospectrograms. Some parts of the spectra
were saturated due to very strong signals.  However, the presented results are
not influenced by these saturated data.

The final frequency range 22--5000 MHz range is complete except a minor
100--150 MHz gap and the sub-range 800--1000 MHz is duplicated. The final
time-frequency radiospectrum was chosen in the time interval 11:55--12:07 UT
(Figure~\ref{fig3}). Because the 2000--5000 MHz Ond\v{r}ejov radiospectrograph
registered also a weak radio emission at about 11:53:54--11:55:00 UT, i.e.,
before the start of all other bursts in the 22--5000 MHz range, these data were
also analyzed.

We also used soft X-ray fluxes observed by the GOES 15
  satellite analyzing the flux data of 2 s cadence for this time interval.
  Moreover, we evaluated photometrically reduced EUV images from the SDO/AIA instrument
  \citep{2012SoPh..275...17L} in several channels
  with the AIA stray-light point-spread function correction applied
  \citep{2013ApJ...765..144P}.

\subsection{Description of radio data}

\begin{figure}
\begin{center}
\includegraphics[width=12cm]{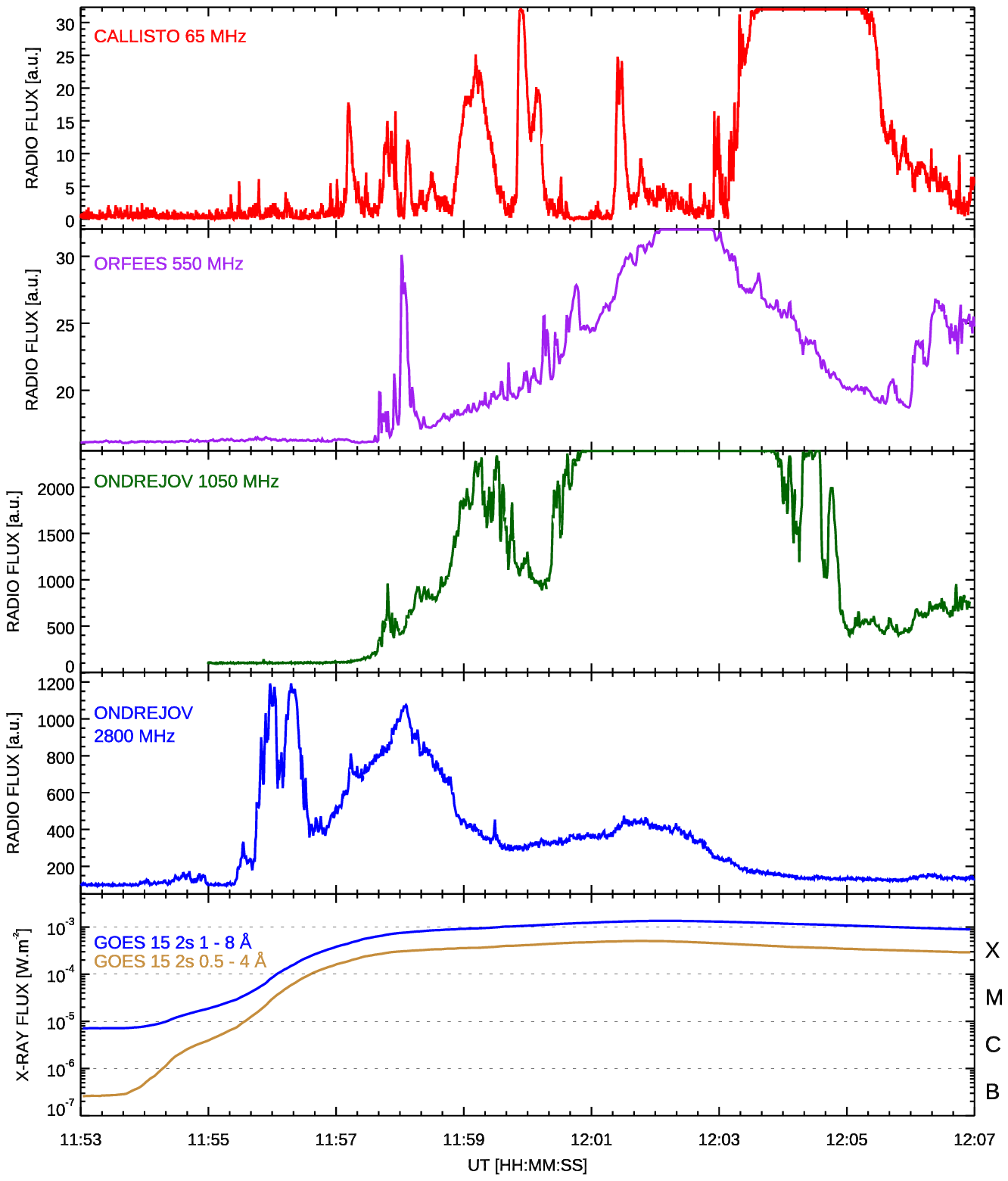}
\end{center}
\caption{Time profiles of the radio flux on 65, 550, 1050, 2800 MHz and GOES 15 0.5-4~\AA~ and 1-8~\AA ~ X-ray fluxes ~ during the 2017 September 6 flare.}
\label{fig1}
\end{figure}

\begin{figure}
\begin{center}
\includegraphics[width=12cm]{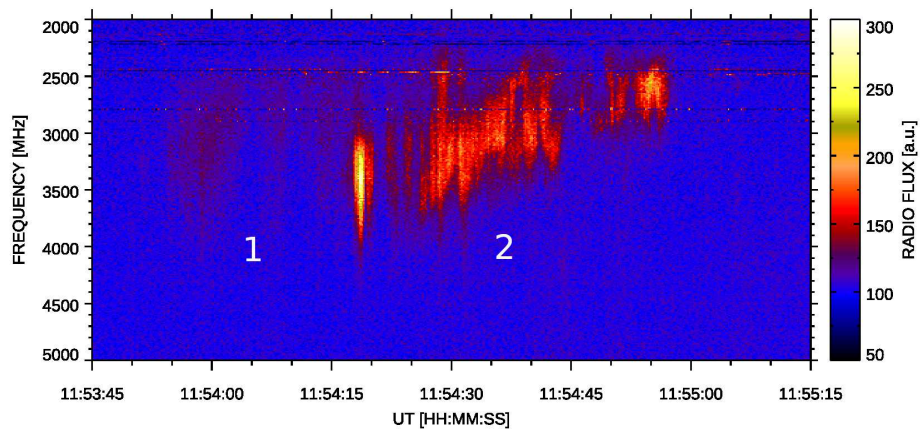}
\end{center}
\caption{The 2000--5000 MHz radio spectrum showing the double DPS with the parts 1
  and 2 in the pre-impulsive flare phase.}
\label{fig2}
\end{figure}

\begin{figure}
\begin{center}
\includegraphics[width=11cm]{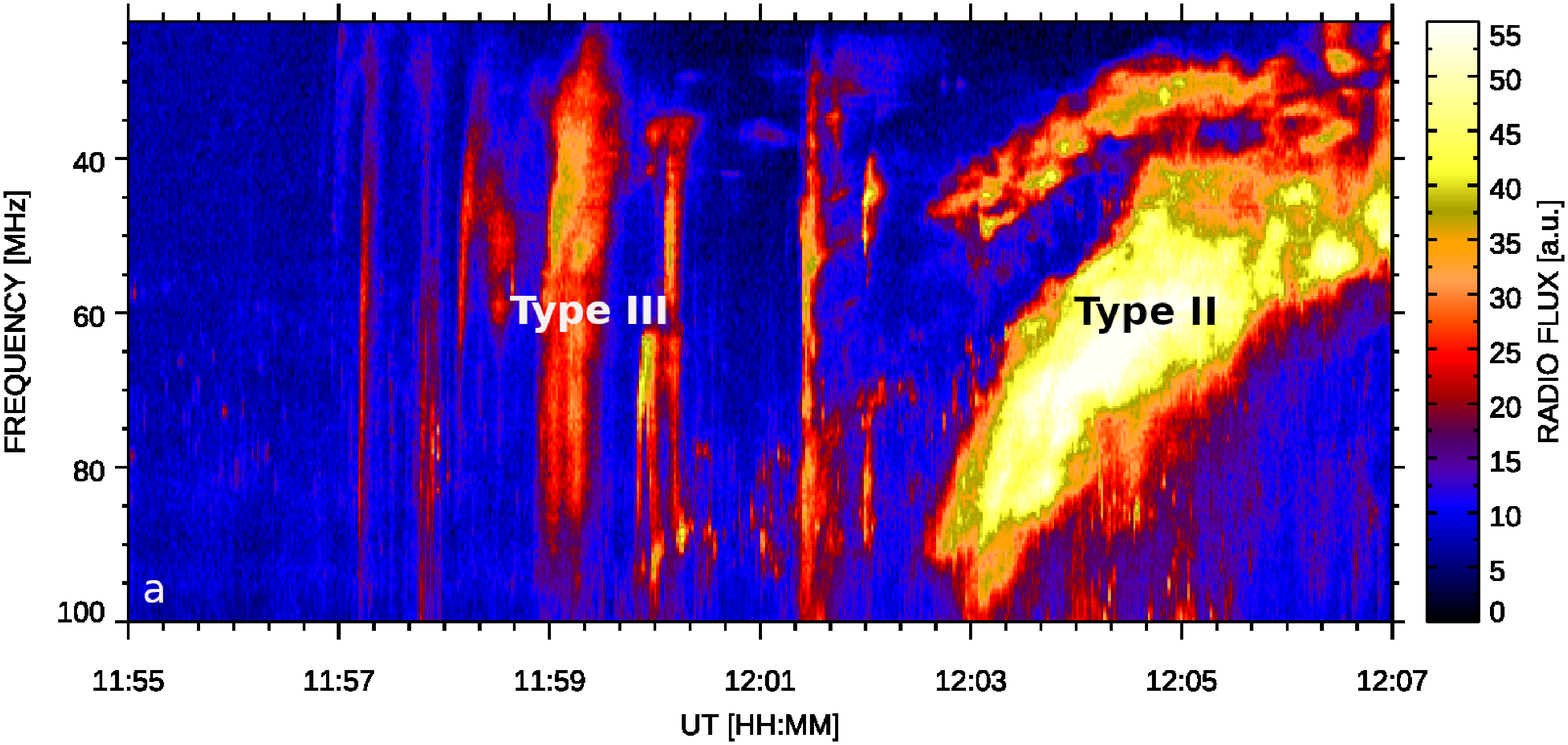}
\includegraphics[width=11cm]{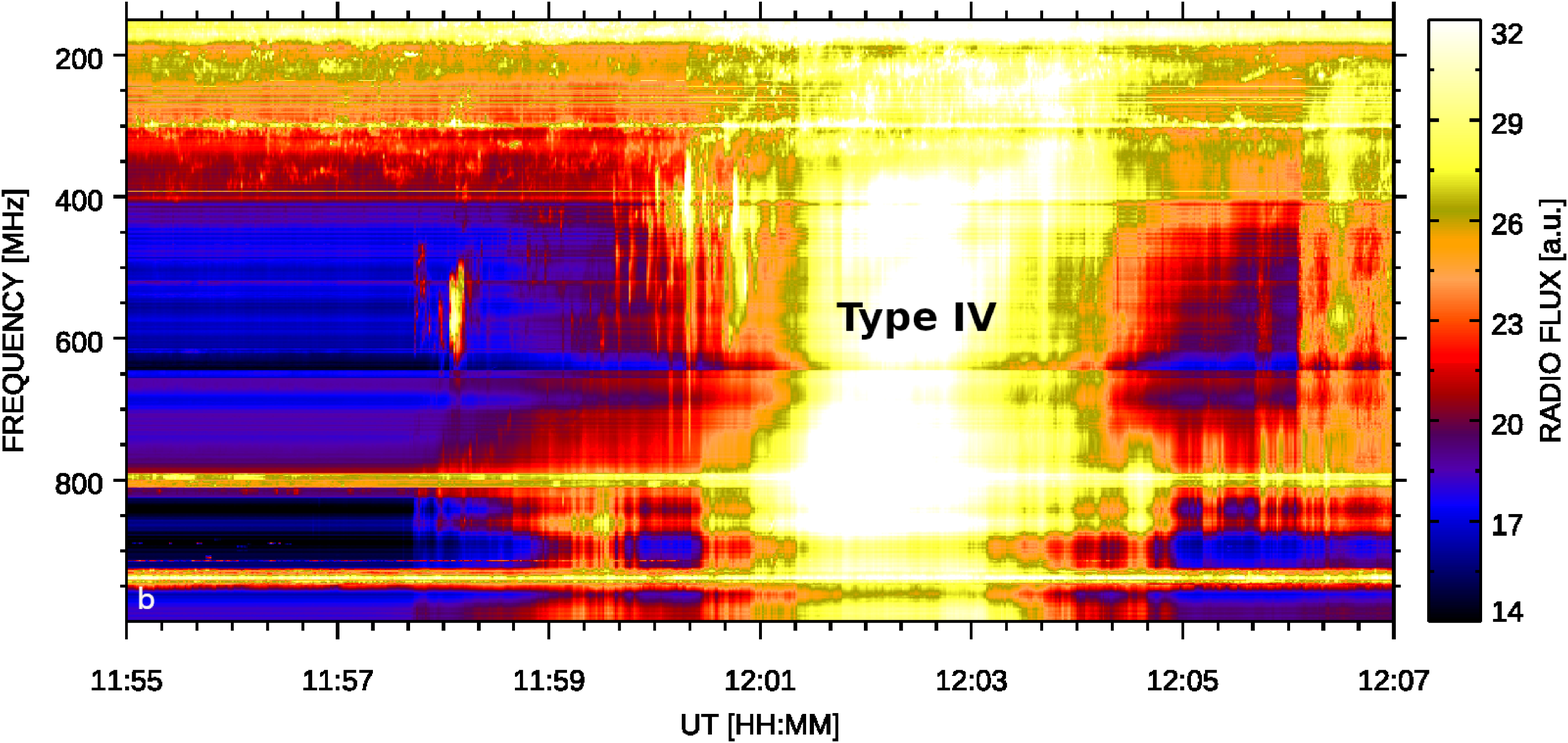}
\includegraphics[width=11cm]{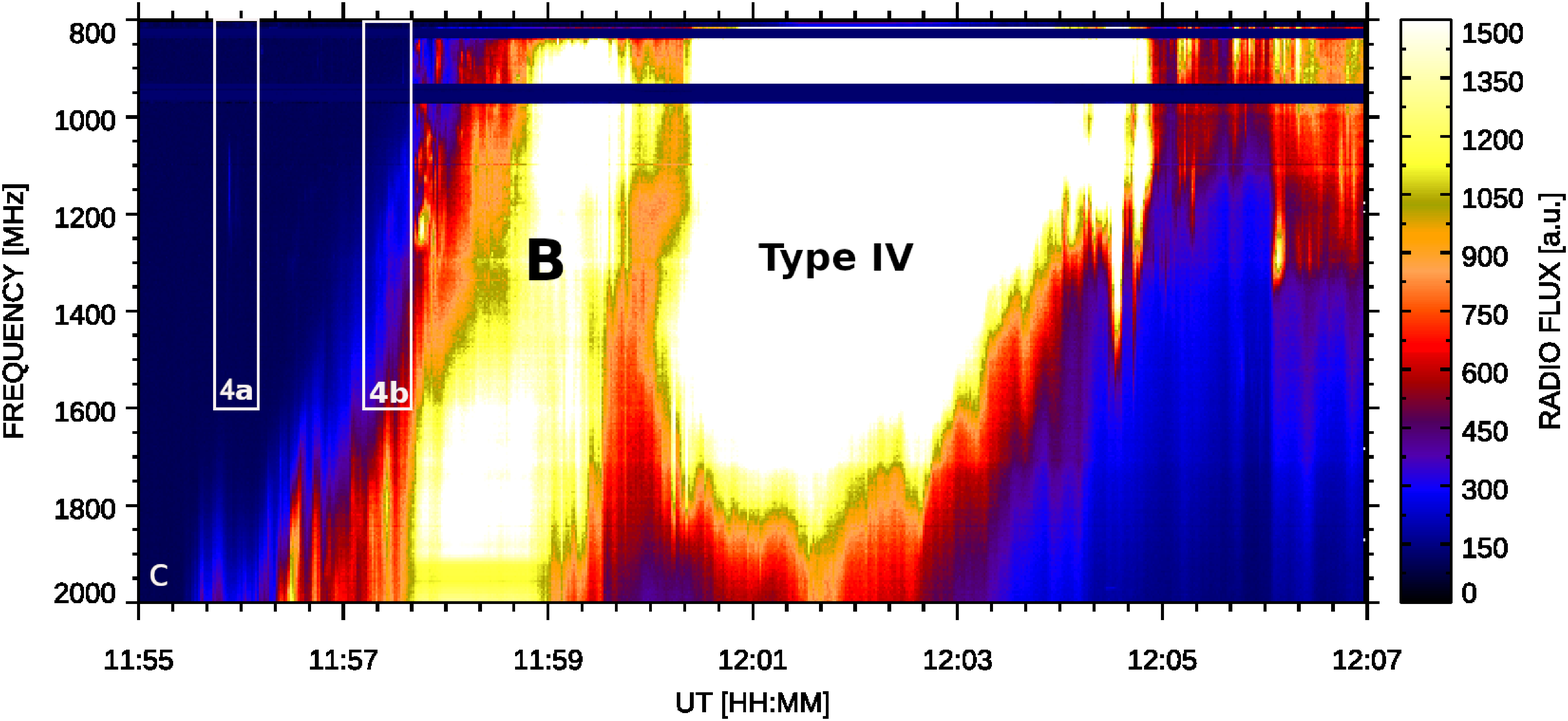}
\includegraphics[width=11cm]{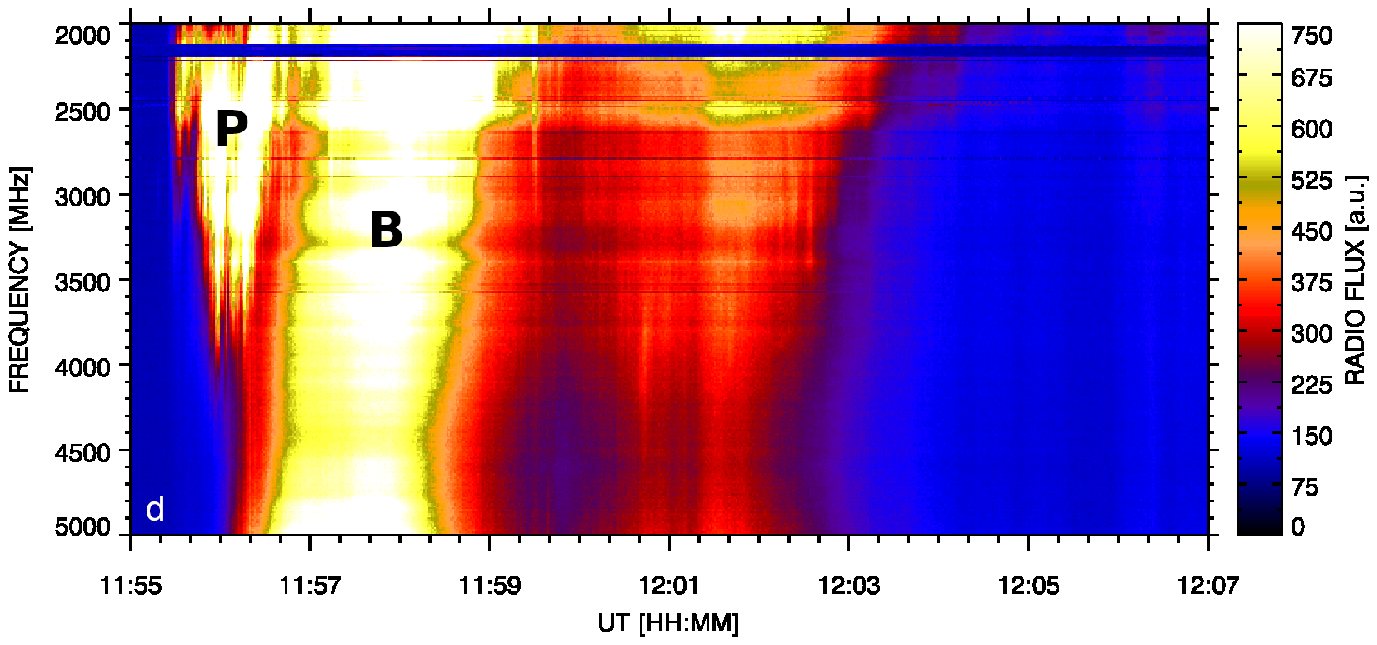}
\end{center}
\caption{The 22--5000 MHz radio spectrum of the 2017 September 6 flare.
  a) The 22--100 MHz Greenland-Callisto spectrum,
  b) the 150--1000 MHz ORFEES spectrum,
  c) the 800--2000 MHz Ond\v{r}ejov spectrum, and
  d) the 2000--5000 MHz Ond\v{r}ejov spectrum.
  P is pulsations and B means the unusual burst drifting from 5000 MHz to 800 MHz
  associated with the type III bursts and followed by the type IV and type II bursts.
  Boxes designated as 4a and 4b in the 800-2000 MHz spectrum show the time-frequency
  regions, where the detailed spectra, presented in Figures~\ref{fig4}a and \ref{fig4}b, are shown.
}
  \label{fig3}
\end{figure}

In Figure~\ref{fig1} the time profiles of the radio flux on 65, 550, 1050 and
2800 MHz together with GOES 15 0.5-4 \AA~ and 1-8~\AA ~ X-ray fluxes during the
2017 September 6 event are shown. Owing to very strong radio fluxes
during this flare some parts of these profiles are saturated. As seen at 2800
MHz and GOES time profiles, the flare started at about 11:54 UT and its maximum
in the radio emission drifts from high to lower frequencies. To understand
better the evolution of the radio emission during this flare,
its radio spectra are presented in Figures~\ref{fig2} and \ref{fig3}.

Figure~\ref{fig2} shows a remarkable drifting pulsation structure (DPS) that
appears on radio waves in the 2200--4200 MHz range just before the main radio
event displayed in Figure~\ref{fig3}. It looks that this DPS consists of two
parts. The part 1 starts at 11:53:54 UT and ends at 11:55:00 UT. The
part 2 is stronger and starts at 11:54:18 UT. It is superposed on the first one
and it drifts from higher to lower frequencies with the mean frequency drift of
about -20 MHz s$^{-1}$. This double DPS is unusual not only that consists of
two parts, but also because it is observed already in the pre-impulsive flare
phase and at high-frequencies. Usually DPSs are observed in the 1--2 GHz range
and they are signatures of the flare impulsive phase
\citep{2015ApJ...799..126N}.

Figure~\ref{fig3} shows a global overview of radio bursts in the
22--5000 MHz range in the time interval 11:55--12:07 UT. There are strong
pulsations (P) at its beginning at 11:55:20--11:57:00 UT in the 1800--4000 MHz
range, see also the details of these quasi-periodic pulsations in panels d and
e in Figure~\ref{fig7}. With some time delay, at about 11:56 UT, the burst
(designated as B) starts at 5000 MHz and drifts to 800 MHz. It reaches the
frequency 800 MHz at about 11:58:30 UT. The mean frequency drift of this burst
B is about -28 MHz s$^{-1}$. In the 150--2000 MHz range this burst is followed
by the type IV radio burst. We note that the burst B is exceptional, mainly due
to its relatively low frequency drift and its form of the emission band lasting
about two minutes and drifting as a whole from 5000 MHz to 800 MHz. Namely, in
this frequency range and in this flare phase we usually observe broadband
continua without any frequency drift or fast drifting bursts having the
frequency drift of about $\pm$ 1 GHz s$^{-1}$ \citep{2001A&A...375..243J}. The
burst B differs from the flare radio bursts in the radio schemes shown in
\cite{1979itsr.book.....K,1985srph.book.....M,1986SoPh..104...19P} and also
from the bursts shown, e.g., in broadband radio spectra in the papers by
\cite{2005ApJ...625.1019P,2014ApJ...782...43B}. During more than 30 years of
observations by the Ond\v{r}ejov radiospectrographs we found only three
examples with the radio bursts that were partly similar to the present burst:
in 1992 February 27, X3.3 flare \citep{1994SoPh..155..171K}, in 1996 July 9,
X2.6 flare \citep{1998A&A...338.1084K}, and in 2000 July 12, X1.9 flare
\citep{2001A&A...369.1104K}. On lower frequencies this burst B is followed by
the type II radio burst at 12:03--12:07 UT in the 22--100 MHz range. Before the
type II burst, at 11:57--12:03 UT in the 22--100 MHz range, type III bursts can
be seen. Moreover, QPPs are omnipresent in the radio spectrum, see e.g.
Figure~\ref{fig6}.

\begin{figure}
\begin{center}
\includegraphics[width=12cm]{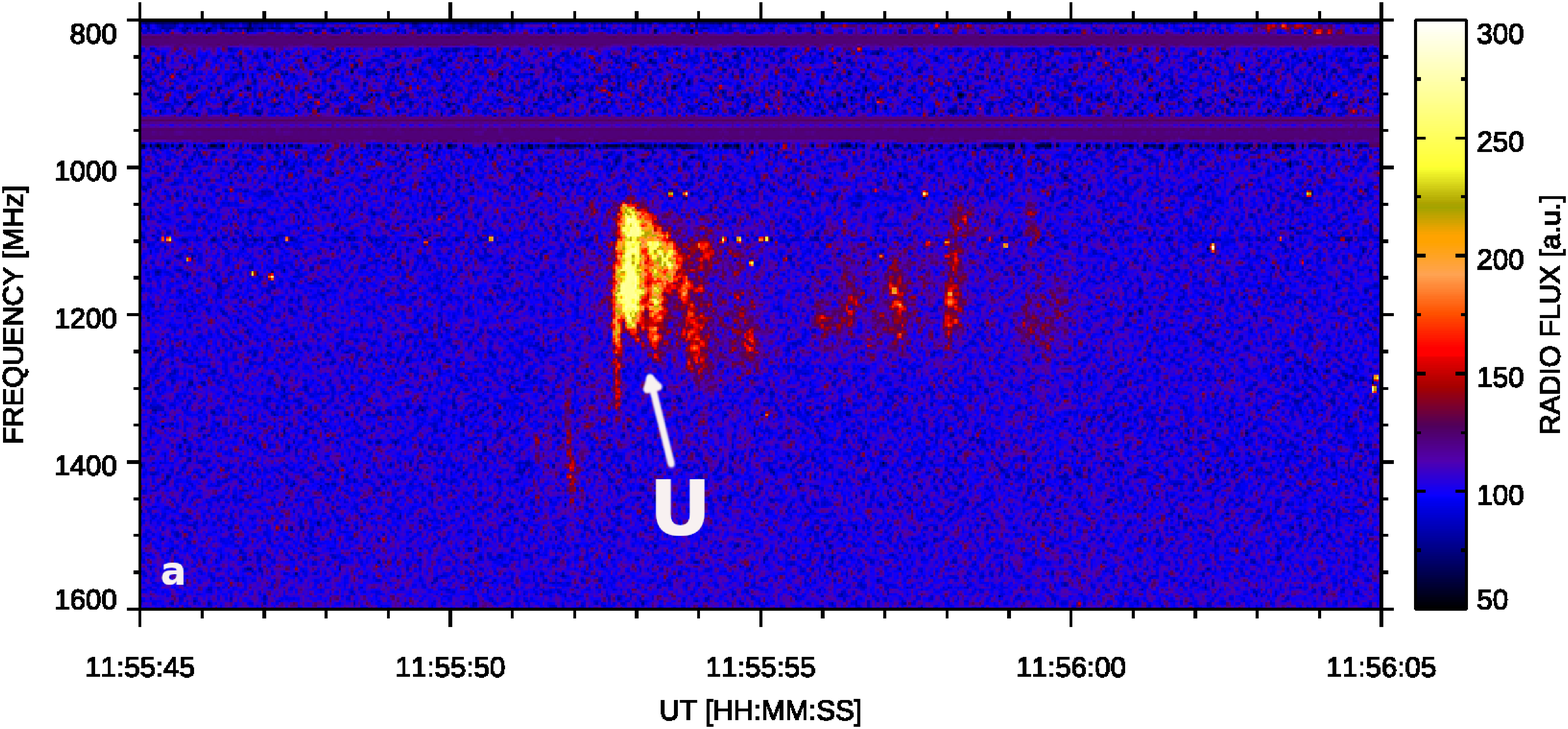}
\includegraphics[width=12cm]{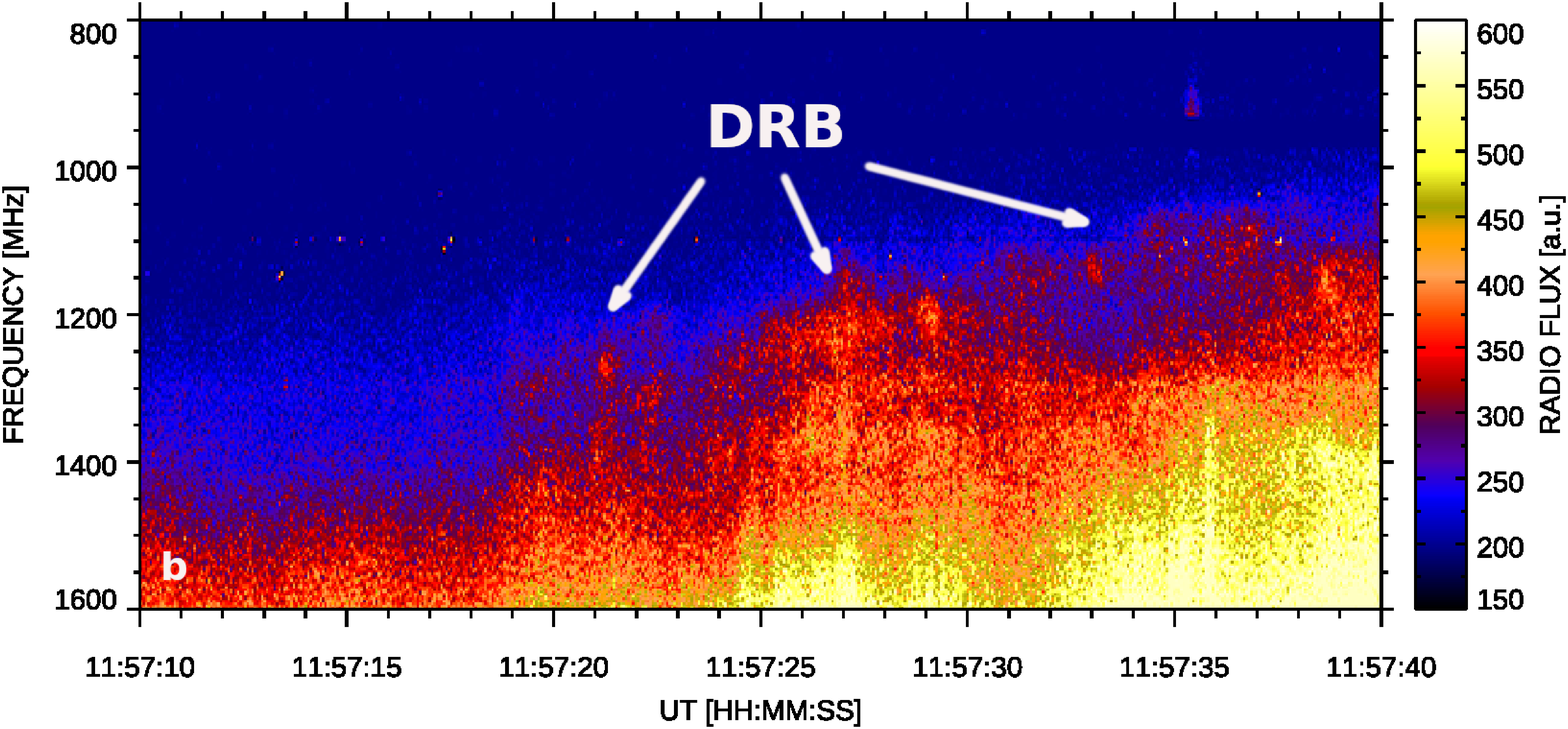}
\end{center}
\caption{Details of the 800--2000 MHz radio spectrum:
  a) Group of bursts starting with the high frequency U-burst
  at 11:55:52--11:56:00 UT in the 1050--1300 MHz range (U), and
  b) drifting radio burst (DRB) at 11:57:19--11:57:37 UT in the 1050--1350 MHz range close to the low-frequency boundary of
  the burst B. see also the boxes 4a and 4b in
  Figure~\ref{fig3}.}
  \label{fig4}
\end{figure}

The 22--5000 MHz radio spectrum is rich in fine structures. The most remarkable
fine structures are shown in Figure~\ref{fig4}. In the panel a of this figure,
in the 1050--1300 MHz range at about 11:55:52--11:56:00 UT, a group of bursts
starting with the high-frequency type U-burst (U) is presented. In the panel b,
in the 1050--1350 MHz range and at 11:57:19--11:57:37 UT, just before the onset
boundary of the drifting burst B, we found an unusual drifting radio burst
(DRB) consisting of many spikes.

\section{Radio spectrum analysis}

\subsection{Periods and phases of radio pulsations}

As mentioned above, pulsations are omnipresent in the radiospectrum of this
flare. Therefore we searched for any periodic signal of these pulsations in the
whole radio spectrum. We used the same technique as presented in
\cite{2017SoPh..292....1K} and \cite{2017SoPh..292...94K}. It is based on the
wavelet transform (WT) \citep{1998BAMS...79...61T} providing a clear detection
of time-frequency evolution of the strong radio wave-patterns. Before
calculations the Ond\v{r}ejov radiospectrum data were resampled to 0.25 s
temporal resolution. No other special data reduction has been applied to the
data of radiospectrographs except a resampling of the ORFEES data to an
equidistant frequency scale. The Morlet mother wavelet, consisting of a complex
sine wave modulated by a Gaussian, was selected to search for radio signal
variability, with the non-dimensional frequency $\omega_0$ satisfying the
admissibility condition \citep{1992AnRFM..24..395F}. The WT was calculated for
the period range starting from 4 time steps with scales sampling the signal
sufficiently as fractional power of two with ${\delta j = 0.4}$ (see Eq. 9.
\cite{1998BAMS...79...61T}). Both the calculated significance of the derived WT
periodicities and the cone-of-influence were taken into account as described in
\cite{2017SoPh..292....1K}. The value of the confidence level was set to
99$\%$.

\begin{figure}
\begin{center}
\includegraphics[width=12cm]{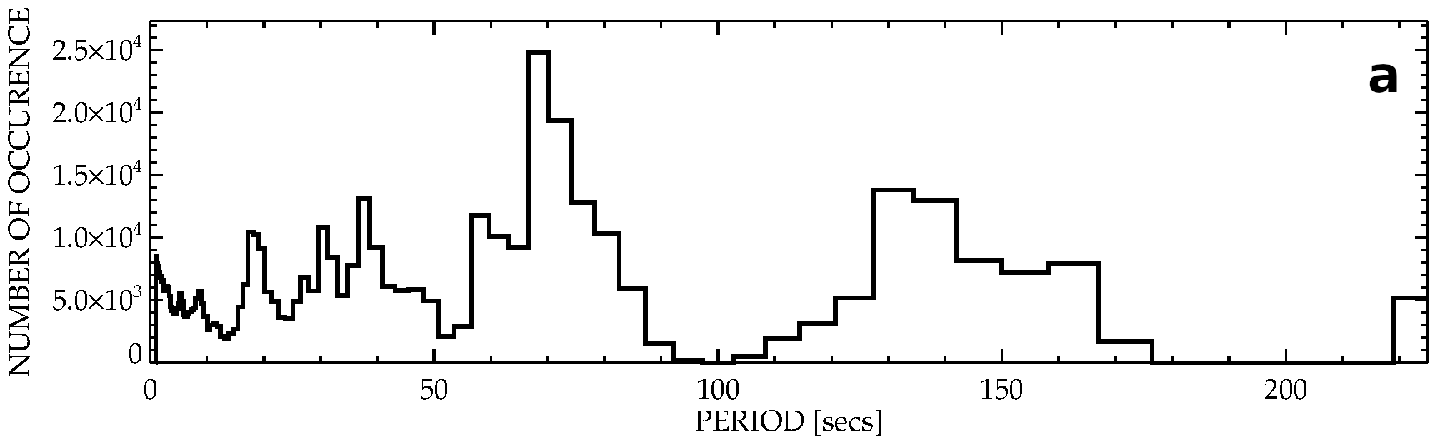}
\includegraphics[width=12cm]{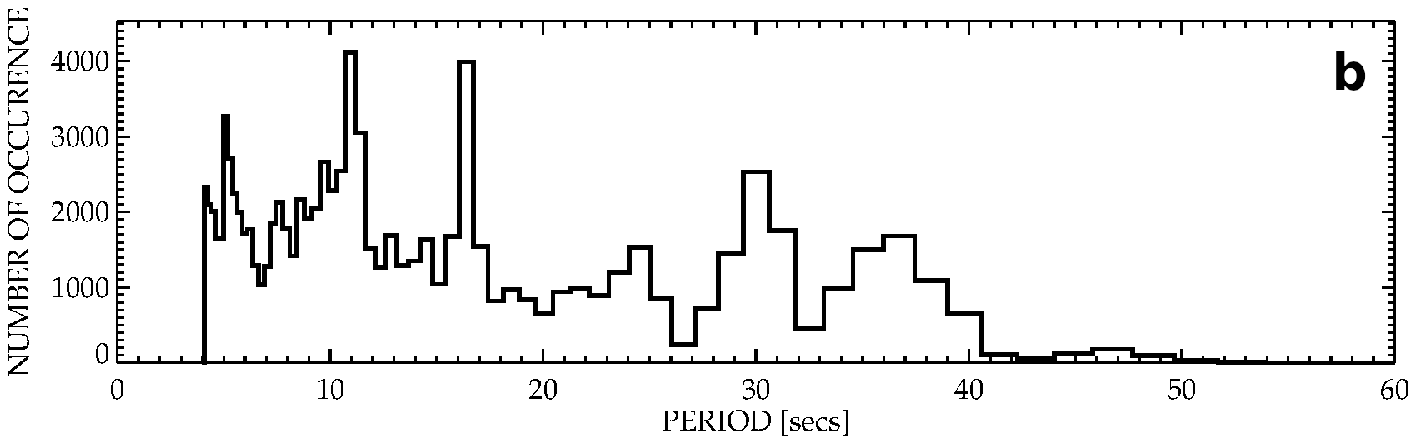}
\includegraphics[width=12cm]{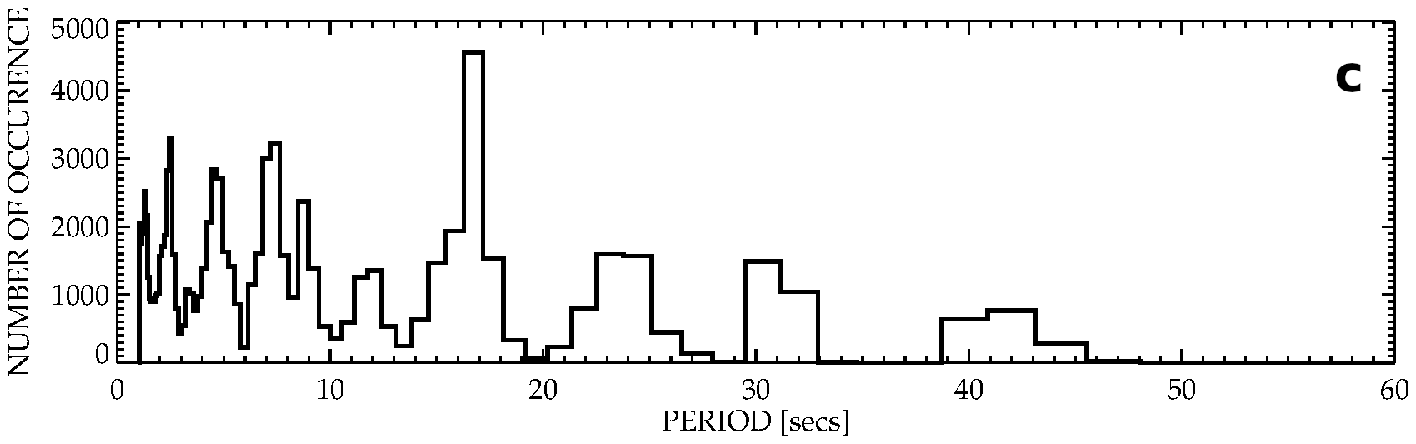}
\includegraphics[width=12cm]{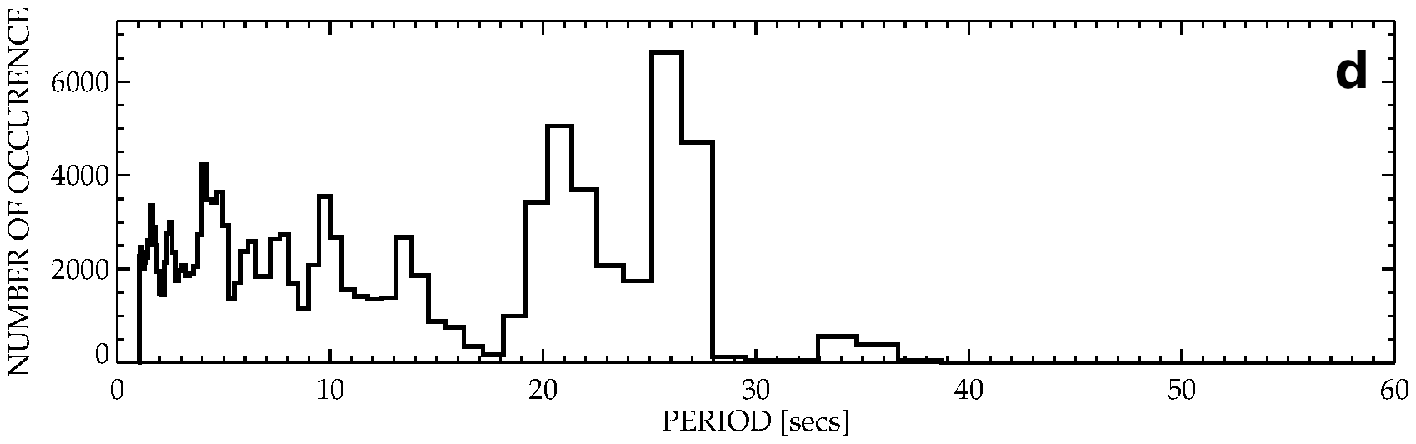}
\end{center}
\caption{Histograms of periods in the radio spectra of
  Callisto-Greenland (22--100 MHz) (a),
  ORFEES (150--1000 MHz) (b),
  Ond\v{r}ejov (800--2000 MHz) (c),
  and Ond\v{r}ejov (2000--5000 MHz) (d).}
  \label{fig5}
\end{figure}

First, we computed the histograms of the detected periods in all four radio
spectra. The results are shown in Figure~\ref{fig5}. Each peak in these
histograms represents a group of significant periodic signals of a roughly
similar period. As seen here, the lower frequency of observations results in a
detection of longer periods.

\begin{figure}
\begin{center}
\includegraphics[width=12cm]{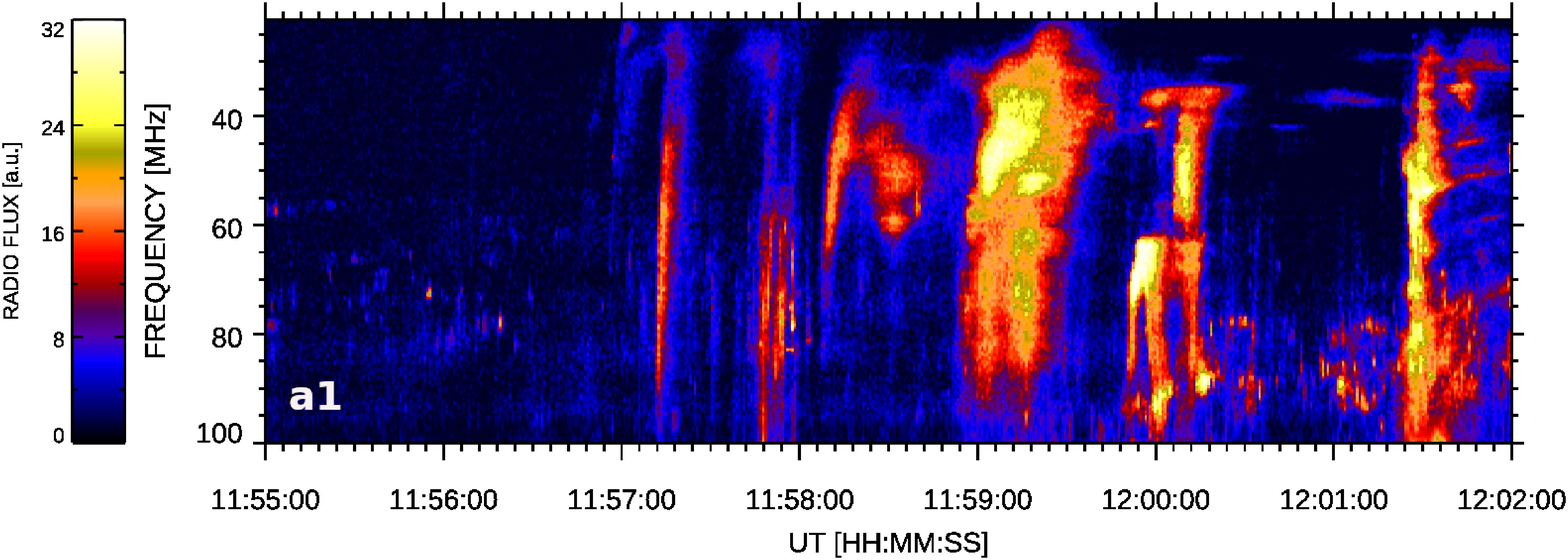}
\includegraphics[width=12cm]{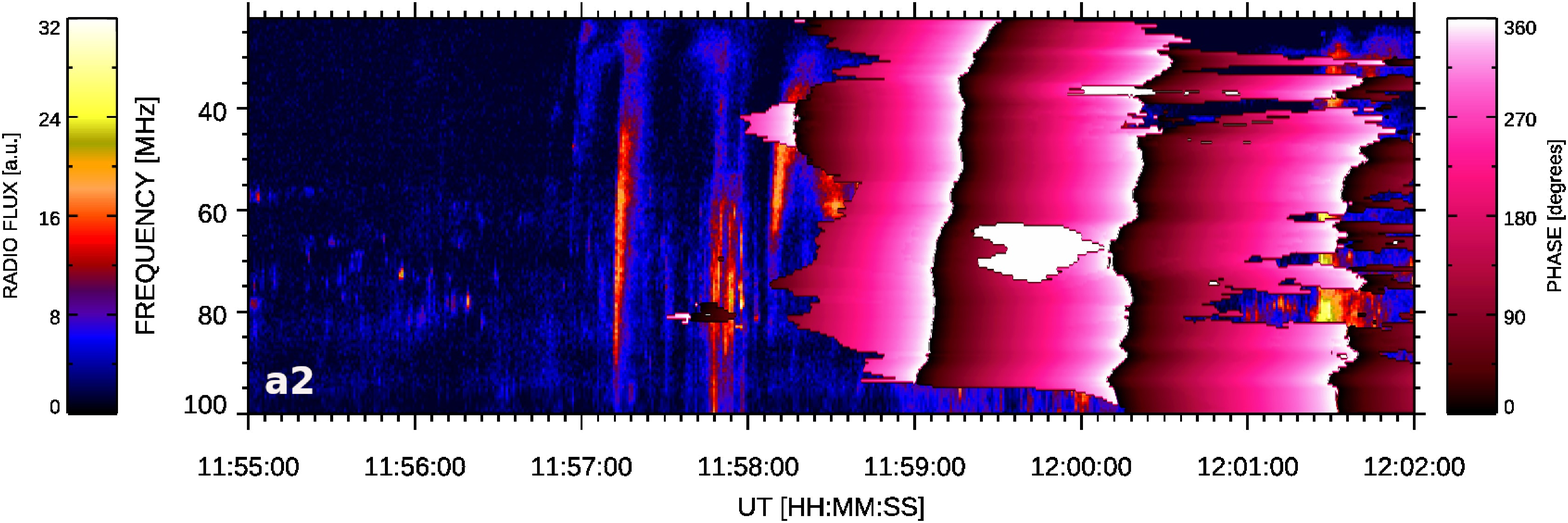}
\includegraphics[width=12cm]{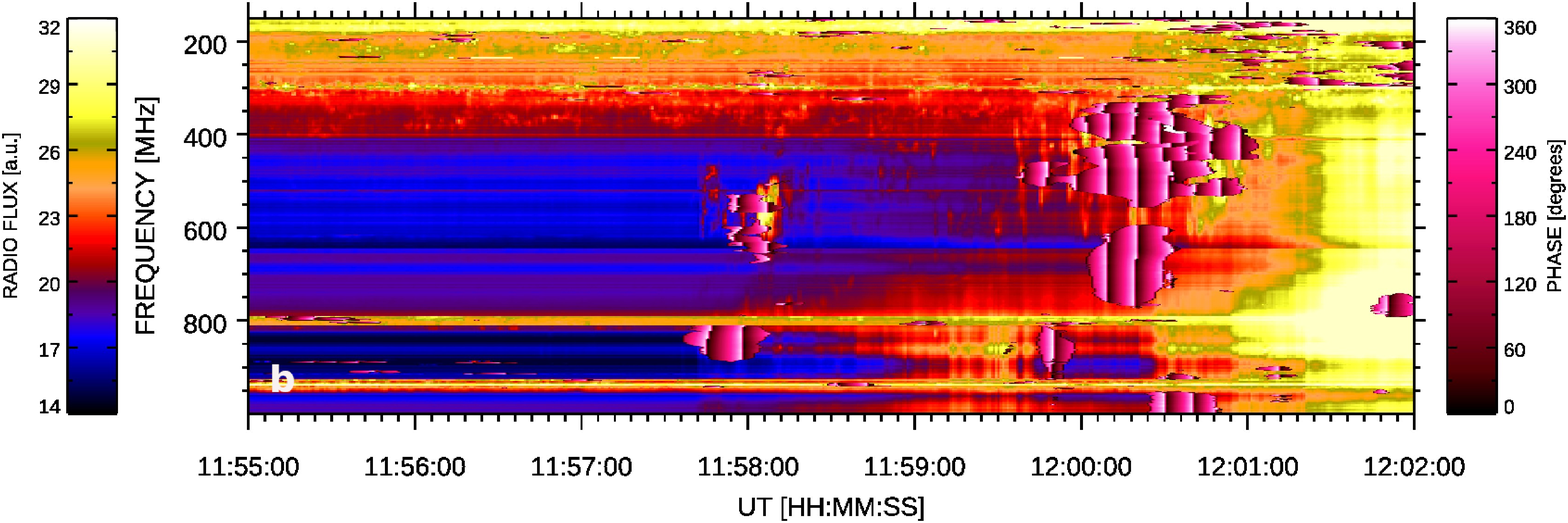}
\includegraphics[width=12cm]{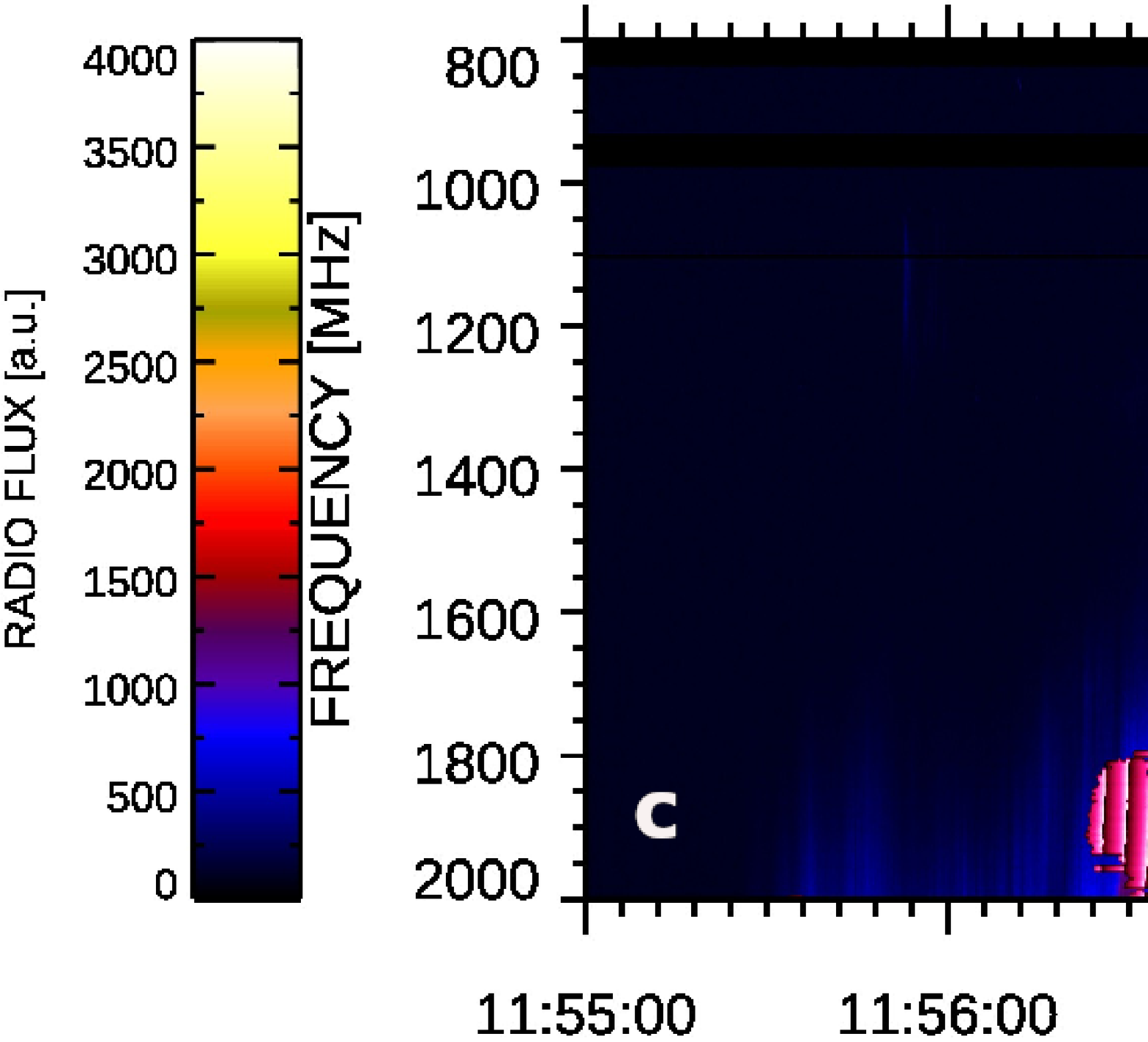}
\includegraphics[width=12cm]{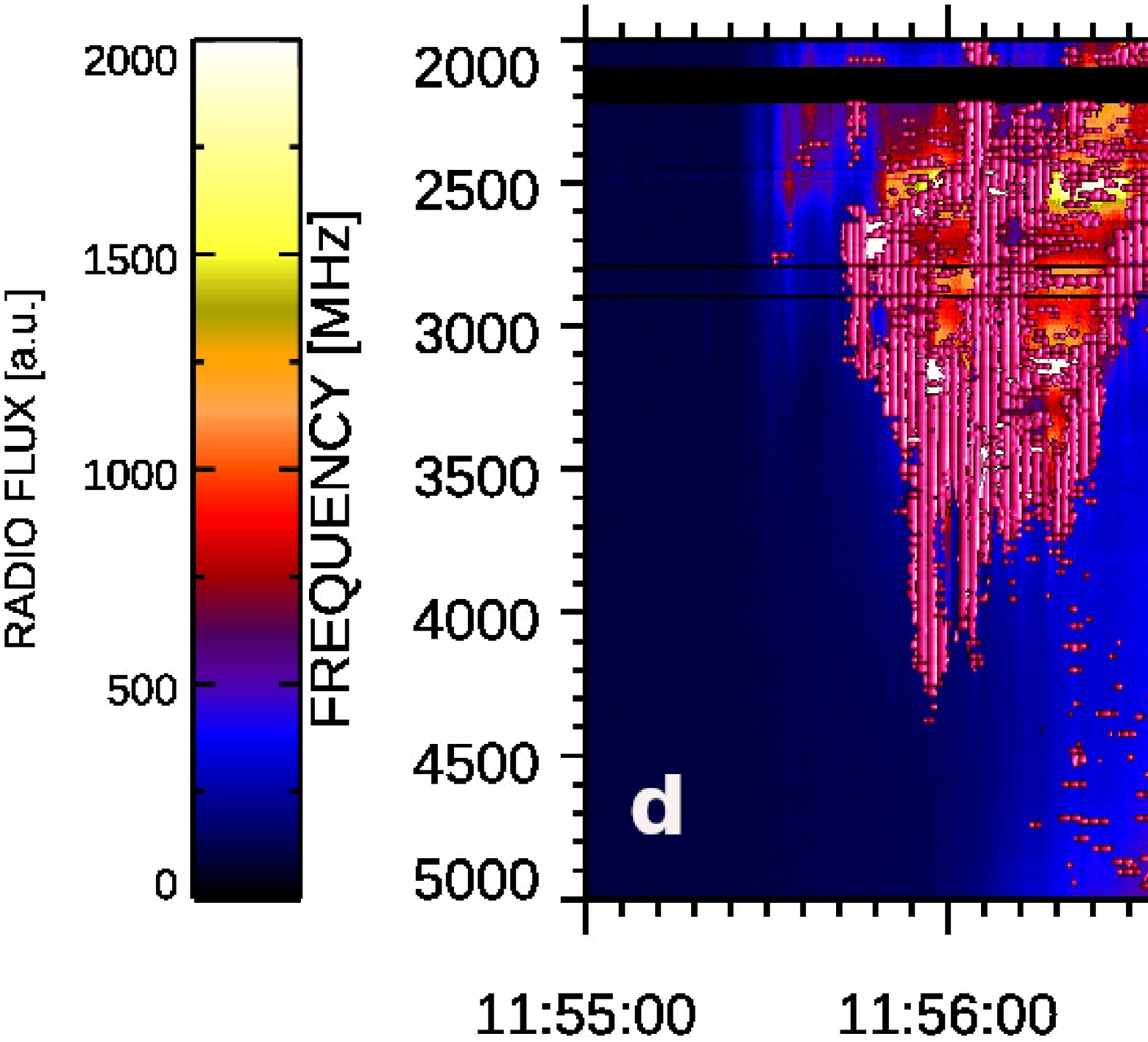}
\end{center}
\caption{a1) The original 22-100 MHz spectrum. a2)--d) Phase maps (pink areas with the black lines showing the zero phase of
  pulsations) overplotted on the radio spectrum for periods 53--100 s in the
  22--100 MHz range (a2), 8--14 s in the 150--1000 MHz range (b), 3--6 s in
  the 800--2000 MHz range (c), and 1--2 s in the 2000--5000 MHz range (d).
  The black part in the 800--2000 MHz spectrum after 12:00:25 UT means
  saturated radio data.}
  \label{fig6}
\end{figure}

To show temporal and frequency location of these periods, for each peak in the
period histograms we made maps of the period phases which we overplotted on the
radio spectra. Because of too many such maps, we present here only examples of
the most interesting maps (Figure ~\ref{fig6}). In the 22-100 MHz range the
phases fully cover the original radio spectrum, therefore the 22-100 MHz
original spectrum is added (Figure ~\ref{fig6}a1). For other maps the original
spectra are much better visible and they can be also seen in Figure~\ref{fig3}.
We found that most of the pulsations are located at the same positions in the
time-frequency domain. It indicates multi-periodic processes at these locations
in the time-frequency domain. These multi-periodic processes were found at
11:55:20--11:57 UT in the 1800--4000 MHz range, at 11:59--12:00:25 UT in the
800--2000 MHz range, at 11:57:30--11:58:20 UT and 11:59:30--12:01 UT in the
150--1000 MHz range, and at 11:57--12:02 UT in the 22--100 MHz range.

\begin{figure}
\begin{center}
\includegraphics[width=12cm]{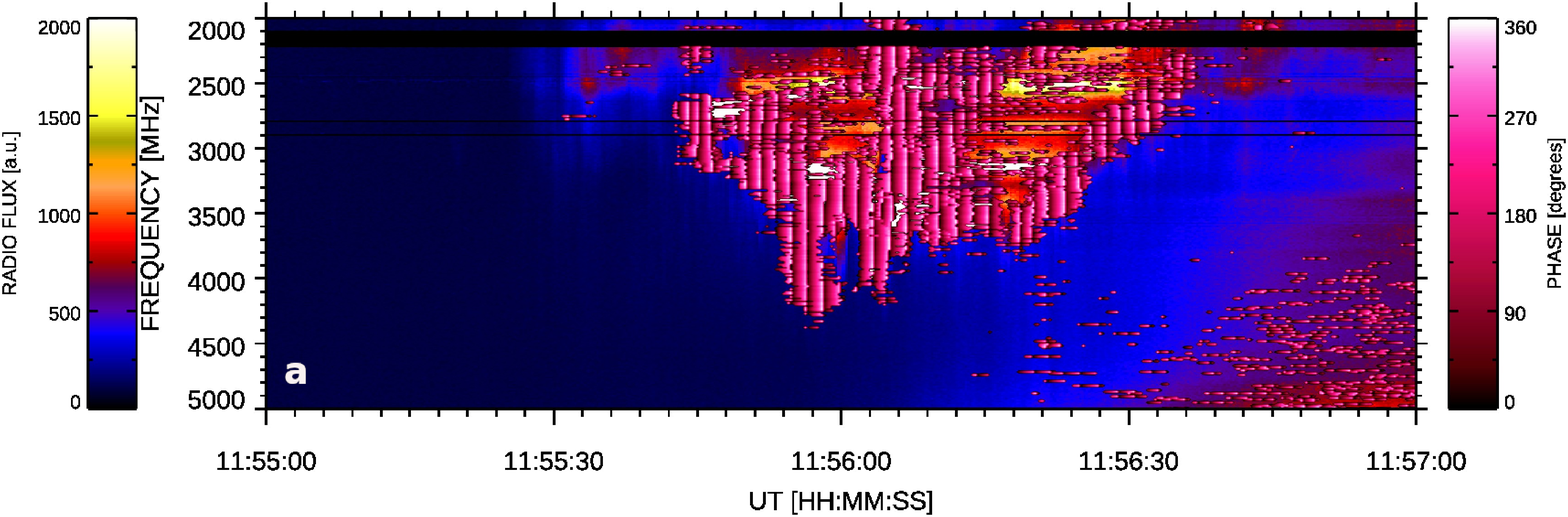}
\includegraphics[width=12cm]{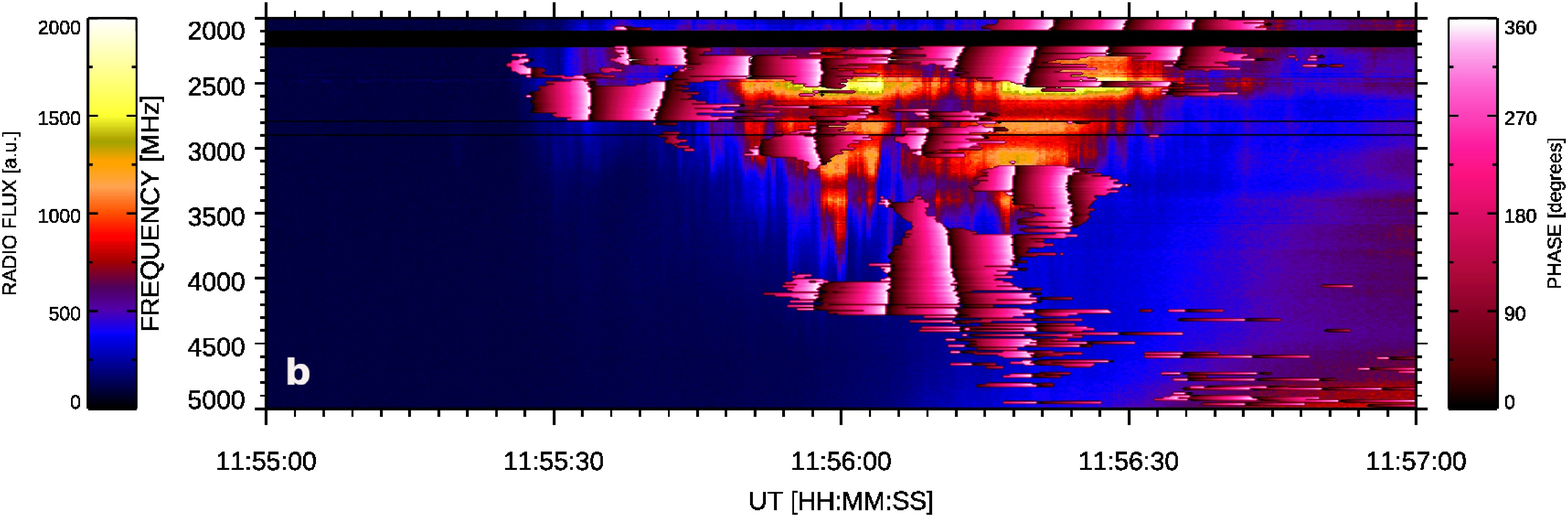}
\includegraphics[width=12cm]{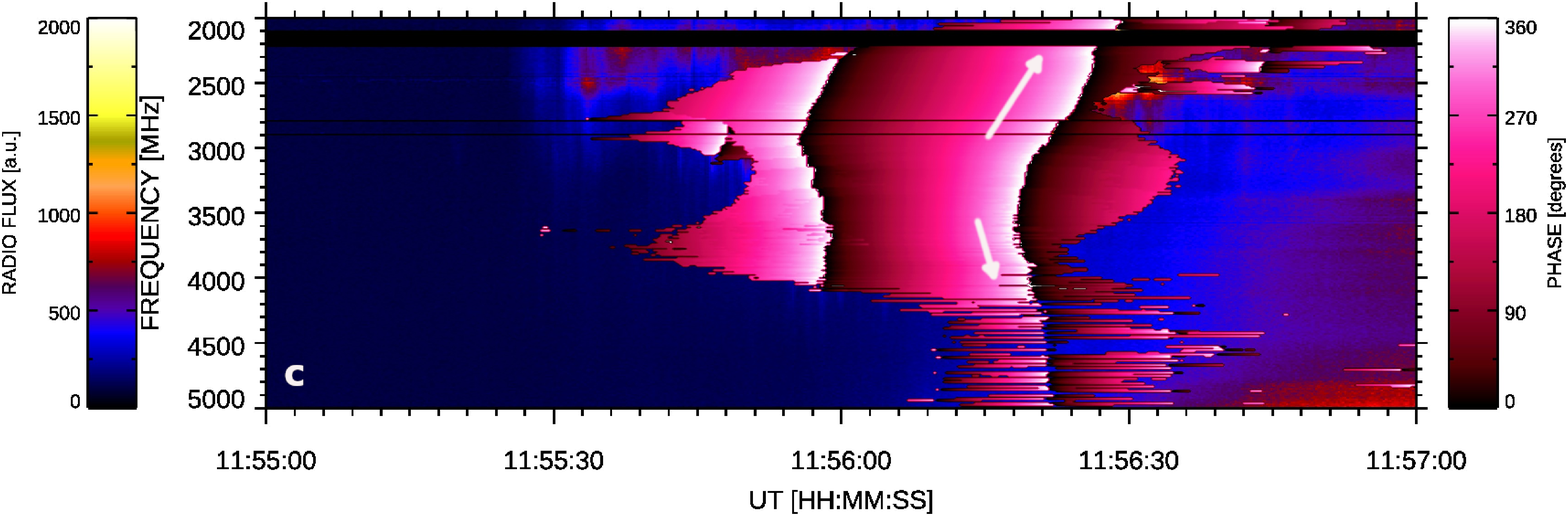}
\includegraphics[width=12cm]{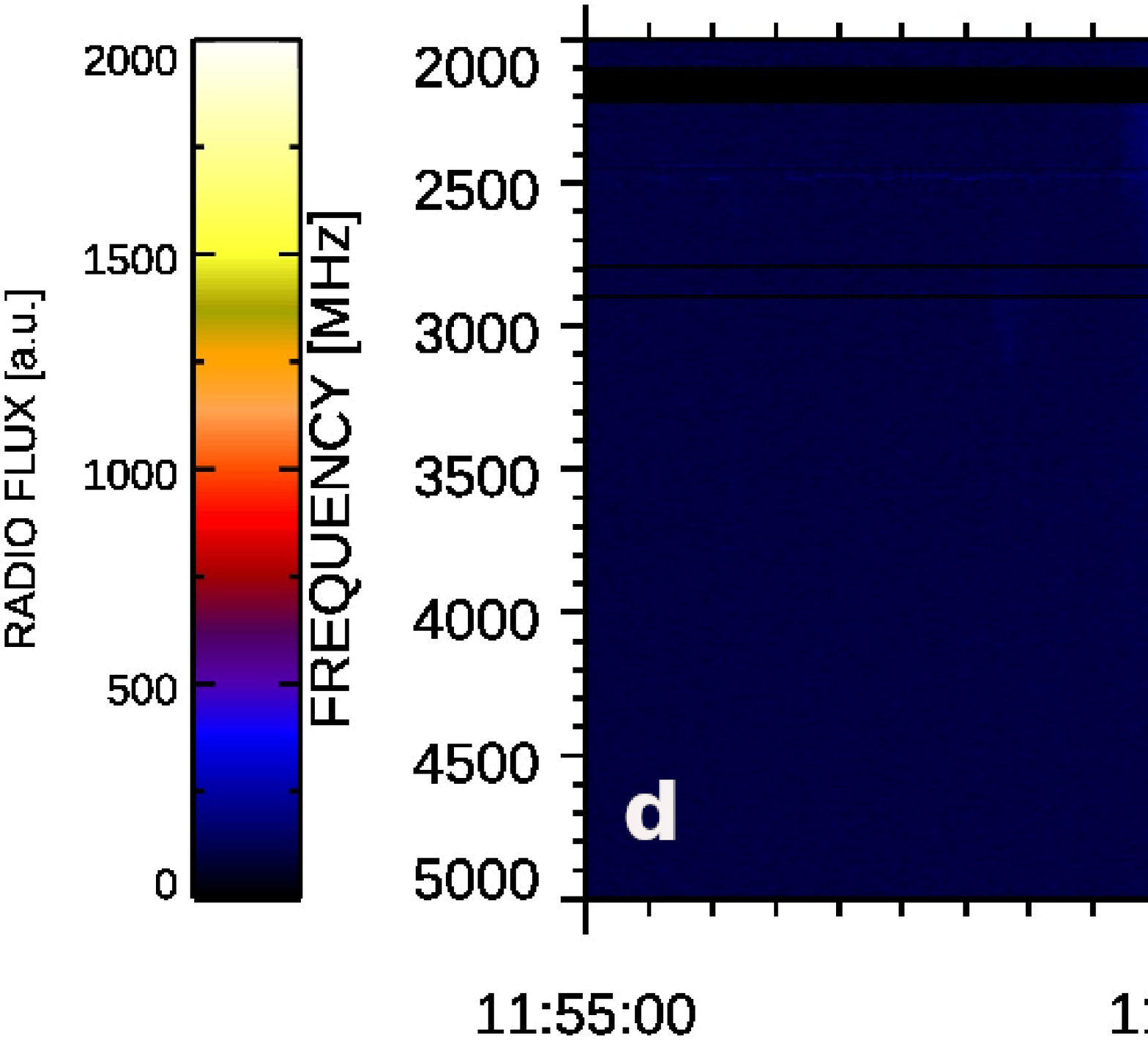}
\includegraphics[width=12cm]{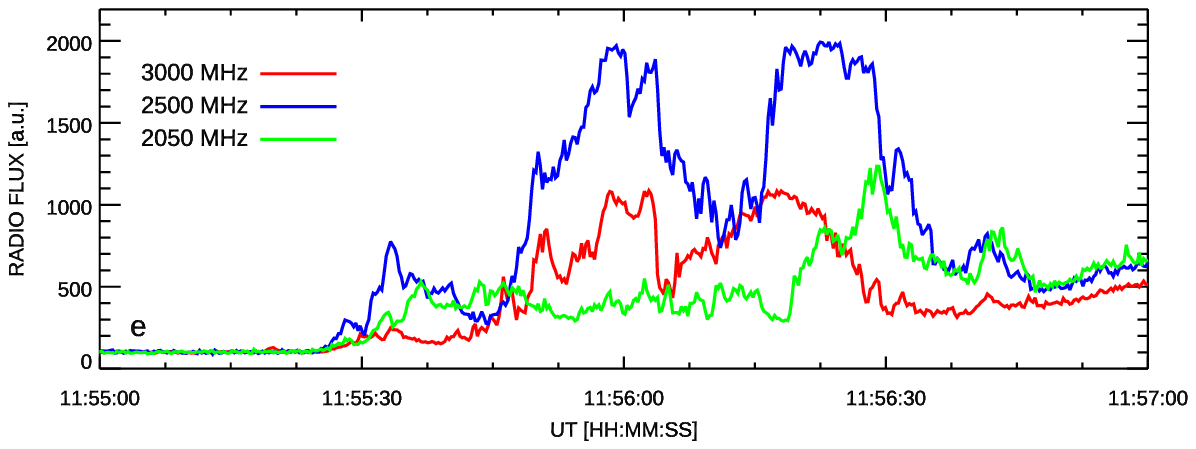}
\end{center}
\caption{Detailed phase maps of pulsations in the 2000--5000 MHz range,
  at time 11:55--11:57 UT, i.e., at time of the first the $\gamma$-ray peak,
  for periods 1--2 s (a),  5.3--8.5 s (b), and 11-30 s (c).
  Arrows in c) show the bi-directional drift of the pulsation phase. d) The original 2000--5000 MHz range spectrum. e) The corresponding
  time profiles of the radio flux on 2050, 2500 and 3000 MHz.}
  \label{fig7}
\end{figure}

\begin{figure}
\begin{center}
\includegraphics[width=12cm]{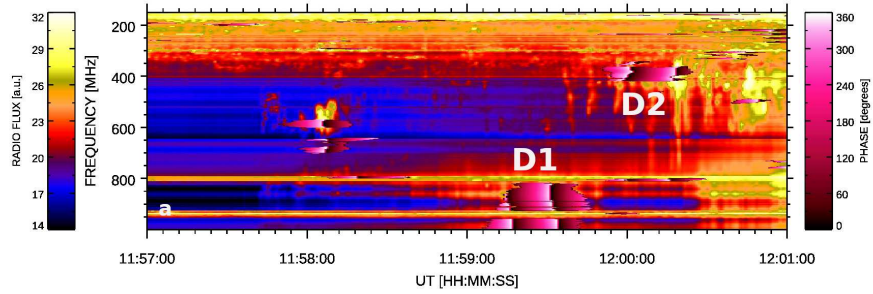}
\end{center}
\caption{Detailed phase map of pulsations at time 11:57--12:01 UT in the
  150--1000 MHz range for periods 15--20 s superposed on the original ORFEES spectrum.
  D1 and D2 designate the pulsations with the phase drifting to higher and lower frequencies, respectively.}
  \label{fig8}
\end{figure}

The most interesting time interval with pulsations is that of the first
$\gamma$-ray peak at 11:56--11:56:50 UT \citep{2019ApJ...877..145L}. Therefore,
in Figure~\ref{fig7} we present three detailed phase maps for the radio
frequency bands 2000--5000 MHz for the periods 1.0--2.0 s, 5.3--8.5 s, and
11--30 s together with the original spectrum and its frequency cuts. As seen
here, the phase of pulsations at 11:55:50--11:56:40 UT with periods 1.0--2.0 s
is synchronized in the whole frequency range for these periods (black lines
that designate the zero phase are vertical (Figure~\ref{fig7}, panel a)). On
the other hand, at about 11:56--11:56:30 UT the pulsations with the periods
5.3--8.5 s and 11--30 s show the frequency drift (Figure~\ref{fig7}, panels b
and c): the negative drift of about -170 MHz s$^{-1}$ for frequencies below
3000 MHz and the positive drift of about 330 MHz s$^{-1}$ for frequencies above
3500 MHz. Furthermore, in Figure~\ref{fig8} we show similar oppositely drifting
pulsations for periods 15--20 s at 11:59:10--11:59:40 UT in the 800--1000 MHz
range (D1) and at 11:59:50--12:00:10 UT in the 350--450 MHz range (D2). The
frequency drift is 30 MHz s$^{-1}$ and -8 MHz s$^{-1}$, respectively.

\subsection{Association of radio bursts with EUV, white-light, X-ray, and $\gamma$-ray observations}

\begin{figure}
\begin{center}
\includegraphics[width=12cm]{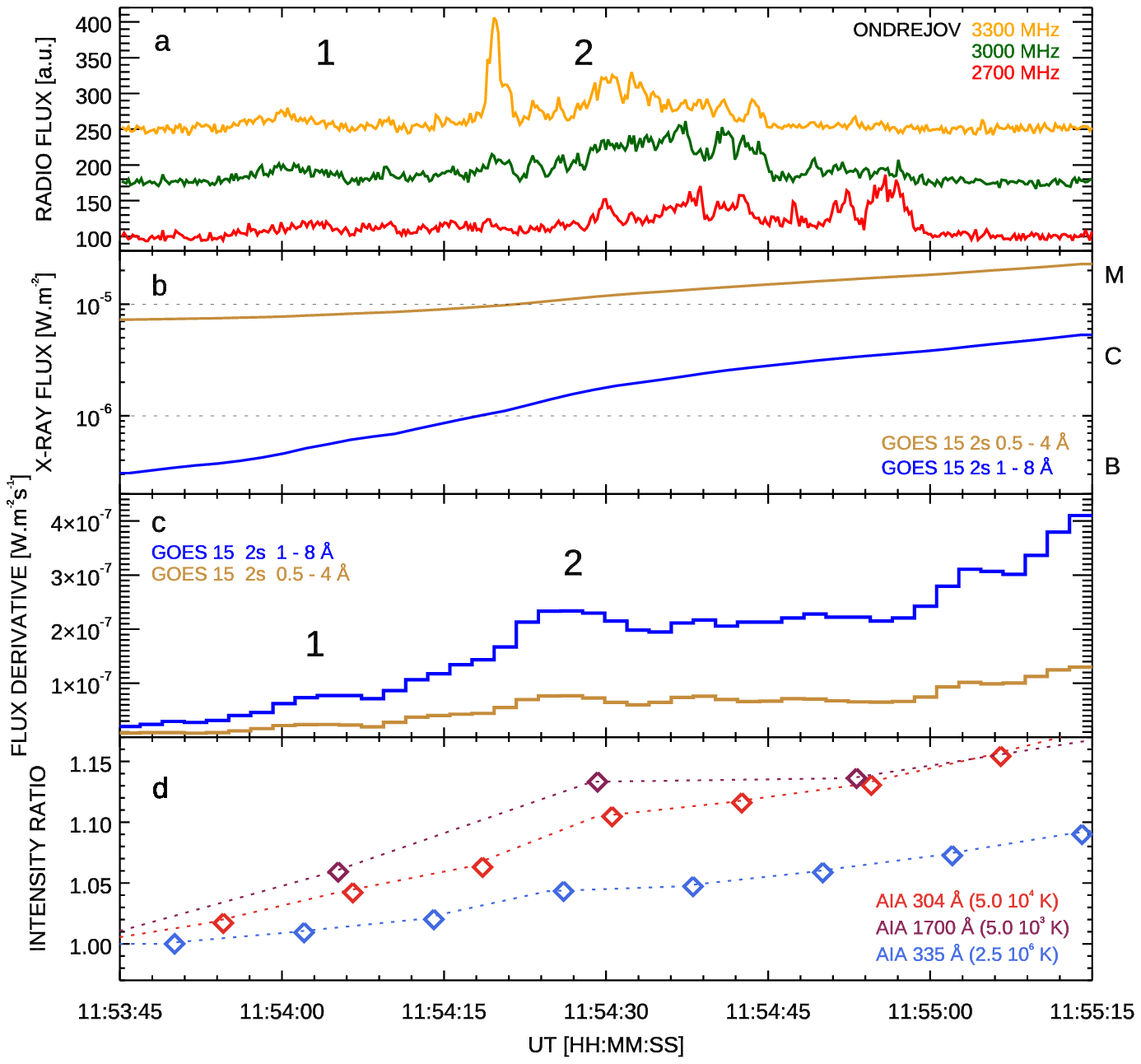}
\includegraphics[width=10cm]{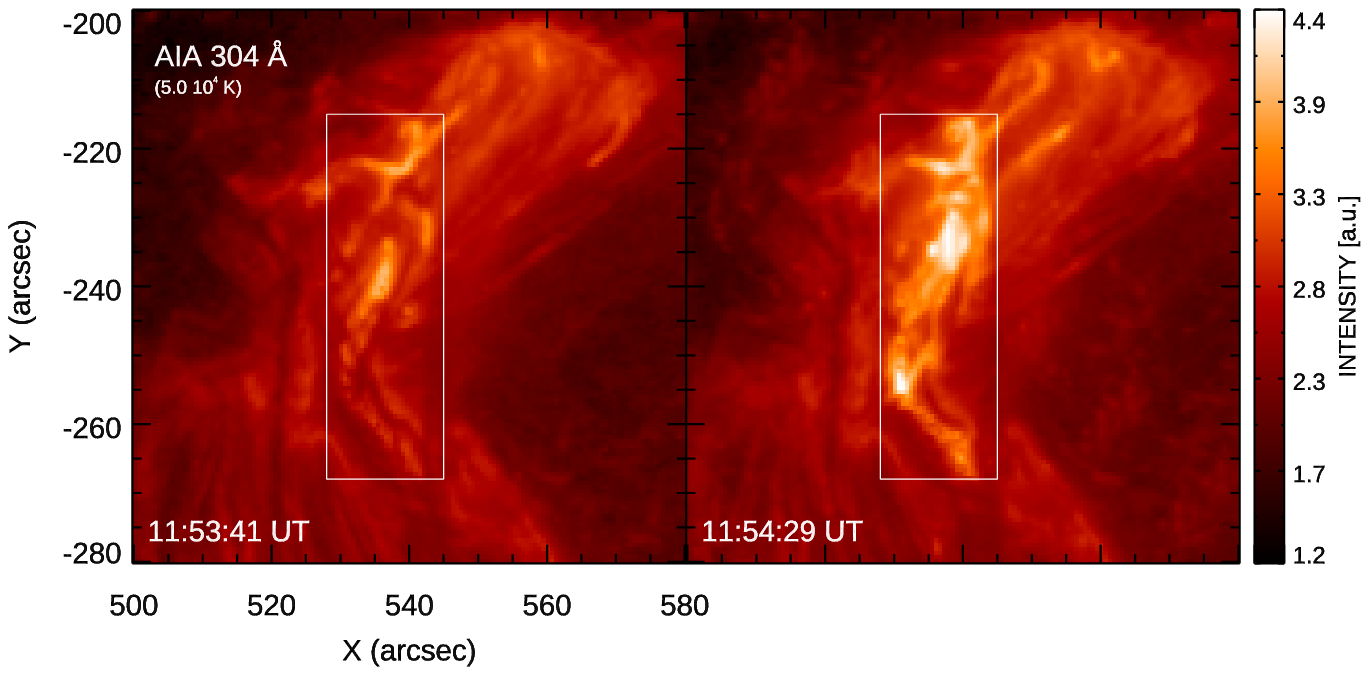}
\end{center}
\caption{Time profiles of the radio flux on 2700, 3000, 3300 MHz (frequency cuts of DPS) (a), GOES 15 0.5-4~\AA~ and 1-8~\AA~fluxes (b),
the time derivative of the GOES 15 fluxes (c),
and the AIA 304, 335 and 1700 \AA~(d) fluxes during the DPS observation. Numbers 1 and 2 in the panel a) mean the part 1 and part 2
of the DPS as designated in Figure~\ref{fig2}. Numbers 1 and 2 in the panel c) mean the first weak and second stronger enhancement.
Bottom panel: The AIA 304 \AA~images
at times 11:53:41 and 11:54:29 UT with the box showing the region where the AIA fluxes were calculated.}
\label{fig9}
\end{figure}

We found several interesting time associations between radio bursts and their
fine structures with the flare phenomena observed in EUV, white-light, X- and
$\gamma$-rays:

a) The initial double DPS was observed at 11:53:54 UT--11:55:00 UT. At this
time interval, in the He\,II 304 \AA\,images (50 kK,
\cite{2012SoPh..275...17L}) several brightenings were observed: at north cross
site at 11:53:53 UT, at south end of Flux rope 2 at 11:54:29 UT, at south end
of Flux rope 1 at 11:54:41 UT, and at south cross site at 11:54:53 UT (see
Figure\,3 in \cite{2018A&A...619A.100H} and Figure~\ref{fig9} bottom panel).
These brightenings were described as caused by interactions between Flux rope 1
and Flux rope 2 \citep{2018A&A...619A.100H}. Furthermore, at about the same
time (at 11:53:20 UT) the first small peak in LYRA Channel 2 (irradiance
measurement in the 1900--2220 \AA~range) was detected
\citep{2018ApJ...867L..24D}.

b) At the time interval 11:55:20--11:57:00 UT in the 1800--4000 MHz range
(Figure~\ref{fig7}) the 1--30 s pulsations with either infinite or with finite
and oppositely drifting phases (see previous sub-section) were observed
simultaneously.
For this time interval the hard X-ray (83--331 keV) and $\gamma$-ray (331--1253 keV)
peaks with contributions from nuclear deexcitation lines,
electron-positron annihilation line at 511 keV, and a neutron capture line at
2.223 MeV along with two components of the bremsstrahlung continuum were
observed \citep{2019ApJ...877..145L}.
At this time also the sunquake \citep{2018ApJ...864...86S} and white-light
flare \citep{2018A&A...620A.183J} started, and oscillations in the $\gamma$-ray
emission with the period $\sim$20 s were detected \citep{2020ApJ...888...53L}.

c) The burst B in Figure~\ref{fig3} started at about 11:56 UT on 5000 MHz and
drifted to lower frequencies down to 800 MHz. Its start coincided with the
start of the upward motion of magnetic ropes at 11:55:53 UT, as can be seen in
Figure 3(a6) in \cite{2018A&A...619A.100H}.

d) Moreover, at the time interval 11:55:52--11:56:00 UT in the 1000--1300 MHz
range, we found a group of bursts starting with the high-frequency type
U-burst. With some short time delay at 11:56:08 UT the EUV writhed structure
was observed as it was documented by \cite{2018A&A...619A.100H}.

\section{Interpretation and Discussion}

We have no information about radio bursts spatial positions. Therefore, in the
following interpretations drawn below, we rely on the results derived from the
spatially resolved observations, already presented in the published articles
about this flare as well as on general knowledge about similar phenomena.

The main topics of the paper are the drifting bursts and the drifting QPPs.
Although we cannot exclude for them the models based, e.g., on the
electron-cyclotron maser mechanism \citep{2019NatCo..10.2276C} or the modulated
gyro-synchrotron emission mechanism \citep{2008ApJ...684.1433F}, for the
drifting phenomena (especially bi-directional) the models based on the plasma
emission mechanism are the most probable \citep{1980gbs..bookR....M}.
Therefore, in the following interpretation we will use the models based on the
plasma emission mechanism.

In this case the time-frequency information from the radio spectrum about the
drifting phenomena means the time-height (in solar atmosphere) information
about these bursts. Thus, the frequency drift of the drifting bursts and the
drifting QPPs can be interpreted by motion of agents (particle beams or
waves) in the vertical direction in the gravitationally stratified solar
atmosphere.

For this purpose, we need to use some models of the solar atmosphere. There are
several models, e.g., models by
~\cite{1947MNRAS.107..426A,1961ApJ...133..983N,1962ApJ...135..138M,1974SoPh...37..443P,1999A&A...348..614M,2002SSRv..101....1A}.
However, only the model of \cite{2002SSRv..101....1A} corresponds to high
plasma frequencies, i.e. to low atmospheric heights, that are of our interest
in this paper. Moreover, this model was derived from the radio observations.
That is why we will use the model of \cite{2002SSRv..101....1A}. However, note
that any estimations made by this model are dependent on this model.

The plasma emission mechanism is a two-steps mechanism
\citep{1980gbs..bookR....M}. Firstly, some agent generates the local plasma
(Langmuir) waves and then these waves are transformed by non-linear processes
to the electromagnetic (radio) waves. The frequency of the radio waves
corresponds to the local plasma frequency (fundamental emission branch) or
double plasma frequency (harmonic emission branch). The intensity of the
emission in the emission source on the fundamental frequency is usually
stronger than that on the harmonic frequency. On the other hand, the emission
on the fundamental frequency is more strongly absorbed than that on
the harmonic frequency, except the case with the fibrous medium
(ducting propagation) \citep{1993ASSL..184.....B}. Because in the emission
source the medium can be fibrous and because we have no additional information
(e.g. polarization of these bursts), we are not able to recognize if the
emission is on the fundamental or harmonic frequency. Therefore, the following
estimations will be presented in the both variants. We note that in some models
of the radio bursts based on the plasma emission mechanism the upper-hybrid
waves are considered instead of the Langmuir waves, see e.g.
~\cite{1982A&A...105..221S}. In the following, if not expressed explicitly we
use the plasma emission mechanism with the Langmuir waves.

In the pre-impulsive phase of the flare, in the 2200--4200 MHz frequency range,
we present the unusual drifting pulsation structure (DPS). It consists from two
parts. Usually DPSs are observed in the impulsive phase and in the 1000--2000
MHz range ~\citep{2015ApJ...799..126N}. They are interpreted as the radio
emission of superthermal electrons trapped in a plasmoid
\citep{2000A&A...360..715K}. In the present case, the DPS was observed at time
of interactions of two magnetic ropes \citep{2018A&A...619A.100H}. This type of
the interaction between ropes is known; it heats the ropes and accelerates
particles \citep{1996SSRv...77....1S}. To confirm the heating and acceleration
processes in the rope interaction region at the DPS time, in Figure~\ref{fig9}
we show the time profiles on radio frequencies 2700, 3000 and 3300 MHz
(frequency cuts of the DPS), the GOES soft X-ray fluxes with their time
derivatives, and the AIA 304, 335 and 1700 \AA~emission fluxes from the rope
interaction region. Comparing the profiles of the time derivative of the GOES
flux (panel c) with the DPS radio profiles (panel a) and the DPS radio spectrum
in Figure~\ref{fig2}, it can be seen that the small enhancement in the GOES
time derivative at 11:54:03 -- 11:54:07 UT (1 in panel c) roughly corresponds
to the weaker part of the DPS (Part 1). On the other hand, the stronger
enhancement in the GOES time derivative at 11:54:24 -- 11:54:30 UT (2 in panel
c) together with those in AIA profiles roughly correspond to the stronger part
of the DPS (Part 2). While the enhancements in the AIA emission fluxes from the
rope interaction region indicate the plasma heating, the enhancements in the
GOES derivatives show a presence of the superthermal electrons as follows from
the Neupert effect \citep{1968ApJ...153L..59N,1993SoPh..146..177D}. These
superthermal electrons can be then trapped in these interacting ropes and
generate by the plasma emission mechanism two DPSs (Part 1 and Part 2 in
Figure~\ref{fig2}). In this interpretation the similar frequency band of the
both parts of DPS means the similar plasma density in these interacting ropes:
5.9 $\times$ 10$^{10}$ - 2.2 $\times$ 10$^{11}$ cm$^{-3}$ (fundamental
emission) or 1.5 $\times$ 10$^{10}$ - 5.4 $\times$ 10$^{10}$ cm$^{-3}$
(harmonic emission). This double DPS resembles to the double structure of the
DPS in the 2017 September 10 flare \citep{2020ApJ...889...72K}, where the
plasma density in the magnetic rope and flare arcade were different and thus
two isolated DPS structures were seen in the radio spectrum.

In the impulsive phase, a negatively drifting burst from 5000 MHz to 800 MHz
was found (burst B in Figure~\ref{fig3}). As shown above this burst is
exceptional. In agreement with EUV observations \citep{2018A&A...619A.100H} we
interpret this burst as generated by the rising magnetic rope that accelerates
superthermal electrons along its trajectory. Then these electrons generate the
drifting burst by the plasma emission mechanism. In such an interpretation the
frequency drift of this burst corresponds to the velocity of the rising rope.
In the Aschwanden's model of the solar atmosphere, for the drifting burst B the
mean velocity is about 310 km s$^{-1}$ (fundamental emission) or 560 km
s$^{-1}$ (harmonic emission), which roughly corresponds to the velocities of
rope and loops in the image plane ($\sim$ 200 -- 380 km s$^{-1}$) shown in
Figure 5 in the paper by \cite{2018A&A...619A.100H}.

Just before and during this burst (burst B), at the time interval
11:55:20--11:57:00 UT in the 1800--4000 MHz we recognized pulsations. At this
time the $\gamma$-ray emission together with the white-light flare
\citep{2018A&A...620A.183J} and start of the sunquake
\citep{2018ApJ...864...86S} were detected. These radio pulsations were
multi-periodic. The pulsations of the shortest periods (1--2 s) were without
measurable frequency drift. On the other hand, the pulsations with the periods
5.3--8.5 s and 11--30 s show the bi-directional drift: the negative frequency
drift for the frequencies below 3000 MHz and the positive drift above 3500 MHz.
The frequency drift of these bi-directional QPPs is much smaller than the drift
of type III bursts, generated by particle beams, in this frequency range.
Namely, assuming the Aschwanden's density model, the frequency drifts of these
pulsations correspond to the velocities 880 km s$^{-1}$ downwards and 1070 km
s$^{-1}$ upwards (fundamental emission) or 1570 km s$^{-1}$ downwards and 1900
km s$^{-1}$ upwards (harmonic emission) in the solar atmosphere. We note that
these velocities are only those in the direction of the density gradient and
the real velocities of the agents generating these QPPs  can be higher.
Considering these velocities and the QPP  detected in the radio spectrum of the
2010 August 1 flare \citep{2017SoPh..292....1K}, where the fast mode
magnetosonic wave was observed \citep{2011ApJ...736L..13L} we propose that the
present bi-directional QPPs are generated by modulation of the radio emission
by fast mode magnetosonic waves. They propagate upwards and downwards from
their source. Thus, their source should be at the altitude in the solar
atmosphere where the plasma frequency is 3000--3500 MHz. In the density model
of the solar atmosphere, the corresponding source altitude and plasma density
is 13800--15700 km and $1.1 \times 10^{11}$ -- $1.5 \times 10^{11}$ cm$^{-3}$
(fundamental emission) or 24800--28200 km and $2.8 \times 10^{10}$ -- $3.8
\times 10^{10}$ cm$^{-3}$ (harmonic emission). Because the quasi-periodic
magnetic reconnection is a natural source of fast mode magnetosonic waves
\citep{2017ApJ...847...98J} we think that the source of the present
magnetosonic waves is the magnetic reconnection. On the other hand, the
pulsations with the periods 1--2 s and with no measurable drift can be
generated by the quasi-periodic magnetic reconnection
\citep{2000A&A...360..715K} or by the electron-cyclotron maser mechanism
\citep{1988ApJ...332..466A}. Thus, it is possible that all these QPPs (with the
periods 1--2 s, 5.3--8.5 s and 11--30 s) are generated by a single process,
i.e. by the quasi-periodic magnetic reconnection.

We found a time association between a group of bursts starting with
the U-burst in the 1050--1300 MHz range with the EUV writhed structure
\citep{2018A&A...619A.100H}. We note that non-thermal (radio) processes usually
precede the thermal (heating) ones, what could explain the short delay between
the EUV writhed-structure observations and the preceding in time radio bursts.
According to the theory, the U-burst is generated by the electron beam
propagating along a closed loop, see e.g. \cite{2004psci.book.....A}.
The EUV writhed-structure looks to be a kinked magnetic rope. The
kinked magnetic rope is not only a closed loop, but in this structure electrons
can be accelerated due to the magnetic reconnection between kinked magnetic
field lines; for the modeled and observed kinked magnetic rope see also the
papers by \cite{2010SoPh..266...91K,2010SoPh..266...71K}.  Thus, the conditions
for the U-burst generation (closed loop and electron acceleration) are
fulfilled in this structure. No other U-bursts were observed. If we assume that
the U-burst is generated by the electron beam, propagating along this
structure, through the plasma emission mechanism, then from the turning
(lowest) frequency of the U-burst we can estimate the density at the top of the
writhed structure as $1.36 \times 10^{10}$ cm$^{-3}$ (fundamental emission) or
$3.1 \times 10^{9}$ cm$^{-3}$ (harmonic emission).

The unusual drifting radio burst (DRB) was detected just before the onset
boundary of the drifting  burst B at 11:57:19--11:57:37 UT in the 1050--1350
MHz range (DRB in Figure~\ref{fig4}b). Its frequency drift is similar to that
of the burst B. Thus, the velocity of the possible agent is also similar to
those of the burst B: 340 km s$^{-1}$ (fundamental emission) or 620 km s$^{-1}$
(harmonic emission). (Note that these velocities are the velocities in the
density gradient direction and real velocities can be higher.) The DRB
resembles to the drifting chains of type I observed in the metric frequency
range \citep{1977sns..book.....E}. The most promising model of these chains is
the model by ~\cite{1982A&A...105..221S}, explaining the type-I chains by
weakly super-Alfvenic shocks generated in the front of emerging magnetic flux.
In agreement with this interpretation and similar frequency drift of the DRB
and the burst B we propose that the DRB is generated by the rising magnetic
rope. The magnetic field around the rising rope is structured, thus some part
of the rising rope can propagate through a region with a relatively
lower magnetic field strength. In this region the velocity of the rising rope
overcomes the Alfv\'en velocity and a weak shock is formed. At this shock the
Langmuir waves (as in the type II burst) or upper-hybrid waves
\citep{1982A&A...105..221S} are generated and after their transformation into
the radio waves the DRB is produced. This interpretation enables to determine
not only the plasma density, but also the magnetic field strength. Assuming the
weakly super-Alfvenic shock (Alfv\'en Mach number $\sim$1) and the plasma
emission mechanism with the Langmuir waves then the mean plasma density and the
magnetic field strength at the DRB source is 1.8 $\times$ 10$^{10}$cm$^{-3}$
and 22.3 G (fundamental emission) or 4.4 $\times$ 10$^{9}$ cm$^{-3}$ and 20.4 G
(harmonic emission). In the case of the DRB generation by the plasma emission
mechanism with the upper-hybrid waves \citep{1982A&A...105..221S}, the
upper-hybrid waves are generated by the double-plasma resonance instability
\citep{1975SoPh...44..461Z,2018A&A...611A..60B} under the condition
$s\omega_{ce} \approx \omega_{uh} \approx (\omega_{pe}^2 +
\omega_{ce}^2)^{1/2}$, where $\omega_{uh}$, $\omega_{pe}$, and $\omega_{ce}$
are the upper-hybrid, electron plasma and electron-cyclotron frequencies,
respectively, and $s$ is gyro-harmonic number. This gyro-harmonic number is
usually much greater than 1. In such a case the upper-hybrid frequency is close
to the plasma (Langmuir wave) frequency. Thus, the estimated plasma density and
magnetic field in the model with the upper-hybrid waves are close to those
estimated above. The height of the DRB source, in the solar atmosphere, in the
Aschwanden's density model is 31000-38000 km (fundamental emission) and
55000-68000 (harmonic emission). Now, using the relation for the model magnetic
field in the solar atmosphere $B(R) = 0.5 (R/R_{\odot}- 1)^{-1.5}$ G (where $R
= R_{\odot} + h$, $R_{\odot}$ is the solar radius and $h$ is the height in the
solar atmosphere) \citep{1978SoPh...57..279D}, the magnetic field strength at
the heights 31000-38000 km is 53-39 G and at the heights 55000-68000 km is
22-16 G. As seen here, the estimated magnetic field for the harmonic emission
is in a better agreement with the model magnetic field than that for the
fundamental emission. But, it does not exclude that the emission of the DRB is
on the fundamental frequency, because the real values of the magnetic field in
the DRB source region, where the weak super-Alfvenic shock is formed, can be
lower than the mean model magnetic field.

Furthermore, in the time interval 11:59--12:01 UT in the 350--400 MHz and
800--1000 MHz ranges we found the pulsations of the 15--20 s period with the
bi-directional drifts (D1 and D2 in Figure~\ref{fig8}). The drift was 30 MHz
s$^{-1}$ for D1 and -8 MHz s$^{-1}$ for D2. In the density model of the solar
atmosphere \citep{2002SSRv..101....1A} these frequency drifts correspond to the
velocity 1360 km s$^{-1}$ downwards and 1690 km s$^{-1}$ upwards (fundamental
emission) or 2400 km s$^{-1}$ downwards and 3000 km s$^{-1}$ upwards (harmonic
emission) in the solar atmosphere. Such velocities look like
velocities of magnetosonic waves. Because they are bi-directional we expect
that these waves were initiated in the source located at the plasma level with
the plasma frequency in the 400--800 MHz range, i.e., with the plasma densities
$2.0 \times 10^{9}$ -- $7.9 \times 10^{9}$ cm$^{-3}$ (fundamental emission) or
$4.9 \times 10^{8}$ -- $2.0 \times 10^{9}$ cm$^{-3}$ (harmonic emission). In
the density model this frequency range corresponds to the height range
47800--85700 km (fundamental emission) or 85700--153000 km (harmonic emission).
Comparing this height range with that of the source of the pulsations at
11:56--11:56:30 UT (see Figure~\ref{fig7}), we can see that the energy-release
source at 11:59--12:01 UT is at higher altitudes than that at 11:56--11:56:30
UT.

As concerns the quasi-periodic pulsations with the period 53-100 s found in the
22–-100 MHz range, we think that they are related to type III bursts
because they have similar frequency drift as type III bursts in this
frequency range. This is clearly seen comparing panels a1 and a2 of
Figures~\ref{fig6}. These QPPs indicate a presence of some continuum formed
from many weak type III bursts.

\section{Conclusions}

In the paper we present not only some unusual radio bursts and fine structures,
but also their interesting time association with the phenomena observed in EUV,
white-light, X-ray and $\gamma$-ray emissions of the 2017 September 6 flare.
Furthermore, we show significant quasi-periodic pulsations (the periods and
phases) in the time-frequency domain of the analyzed radio spectrum
(11:55--12:07 UT and 22--5000 MHz). Especially, the bi-directional QPPs with
the positive and negative frequency drifts are the most remarkable. We note
that the detection of these drifting QPPs were enabled by our new method for
computation of periods and their phases in radio spectra.

We found the double DPS in the pre-impulsive flare phase coupled with
the EUV brightenings caused by an interaction of two magnetic ropes.
Quasi-periodic pulsations at the beginning of the impulsive flare phase, when
the hard X- and $\gamma$-ray emission, white-light flare and sunquake started,
were multi-periodic. While phases of the short periods (1--2 s) have the
infinite frequency drift, the longer periods (5.3--8.5 and 11--30 s) showed the
oppositely drifting phases. These bi-directional QPPs were interpreted as
caused by the magnetosonic waves.

Furthermore, we presented a group of bursts starting with the U-burst at about
the time of the EUV writhed structure and the unusual burst B drifting
from 5000 to 800 MHz, which we interpreted as caused by the rising magnetic
rope. In front of this burst we found the drifting radio burst (DRB) which we
proposed to be a signature of the weakly super-Alfvenic shock, observed on
unusually high frequencies (1050--1350 MHz) and generated by the rising
magnetic rope.

Considering the plasma emission mechanism for all analyzed drifting bursts and
using the density model of the solar atmosphere we estimated the density, wave
velocities and source heights for the bi-directionally drifting QPPs, density
for the pre-impulsive DPS and U-burst, and density and magnetic field for the
drifting structure. We showed that the energy-release process (magnetic
reconnection) moved upwards in the solar atmosphere during the flare.

\acknowledgements The authors thank the referee for useful comments that
improved the paper. We acknowledge support from the project RVO:67985815 and GA
\v{C}R grants 18-09072S, 19-09489S, 20-09922J, and 20-07908S. This work was
also supported by the Science Grant Agency project VEGA 2/0048/20 (Slovakia).
Help of the Bilateral Mobility Projects SAV-18-01 of the SAS and CAS is
acknowledged. This article was created in the project ITMS No. 26220120029,
based on the supporting operational Research and development program financed
from the European Regional Development Fund. The authors are indebted to the
Institute for Particle Physics and Astrophysics, ETH Zurich and FHNW
Brugg/Windisch (Switzerland) for the Callisto data as well as the Paris
Observatory for the ORFEES data. This research has used of NASA Astrophysics
Data System. The wavelet analysis was performed with the software based on
tools provided by C. Torrence and G.P. Compo at {\tt
http://paos.colorado.edu/research/wavelets}.




\newpage

\newpage

\newpage

\newpage

\newpage

\newpage

\newpage

\end{document}